\documentclass[a4paper,11pt]{article}
\usepackage{jheppub}
\usepackage[T1]{fontenc} % if needed

\usepackage{epstopdf}
\usepackage{mathrsfs}
\usepackage{subfigure,graphicx}

\newcommand{\tev}{{\rm{TeV}}}
\newcommand{\gev}{{\rm{GeV}}}

\newcommand{\sL}{\!\scriptscriptstyle L}
\newcommand{\sR}{\!\scriptscriptstyle R}

\newcommand{\PP}{\mathbb{P}}
\newcommand{\PL}{\PP_{\sL}}
\newcommand{\PR}{\PP_{\sR}}

\newcommand{\beq}{\begin{equation}}
\newcommand{\eeq}{\end{equation}}
\newcommand{\beqa}{\begin{eqnarray}}
\newcommand{\eeqa}{\end{eqnarray}}
\newcommand{\nonr}{\nonumber}
	
\title{Charged Lepton Flavor Violating Processes and Scalar Leptoquark Decay Branching Ratios in the Colored Zee-Babu Model }
\author[a,1]{We-Fu Chang,\note{Corresponding author.}}
\author[a]{Siao-Cing Liou,}
\author[b]{Chi-Fong Wong}
\author[c]{and Fanrong Xu}

\affiliation[a]{Department of Physics, National Tsing Hua University, \\101 Sec.2 KuangFu Rd., Hsinchu 300, Taiwan }
\affiliation[b]{School of Physics and Astronomy, Sun Yat-sen University\\
 No. 135, Xingang Xi Road, Guangzhou, 510275, P. R. China}
\affiliation[c]{Department of Physics, Jinan University\\Guangzhou 510632, P. R. China}

\emailAdd{wfchang@phys.nthu.edu.tw}
\emailAdd{sawout@gmail.com}
\emailAdd{freeman.cf.wong@gmail.com}
\emailAdd{fanrongxu@jnu.edu.cn}

\abstract{
We consider a neutrino mass generating model which employs a scalar leptoquark, $\Delta$, and a scalar diquark, $S$.
The new scalars $\Delta$ and $S$ carry the standard model $SU(3)_c\times SU(2)_L\times U(1)_Y$ quantum numbers $(3,1,-1/3)$ and $(6,1,-2/3)$, respectively.
The neutrino masses are generated at the two-loop level, as in the Zee-Babu model\cite{Zee-Babu}, and $\Delta/S$ plays the role
of the doubly/singly charged scalar in the Zee-Babu model.
With a moderate working assumption  that the magnitudes of the six Yukawa couplings between $S$ and the down-type quarks are of the same order, strong connections  are found between the neutrino masses and the charged lepton flavor violating processes.
In particular, we study $Z\rightarrow \overline{l} l'$, and $l\rightarrow l' \gamma$  and  find that some portions of the parameter space of this model are within the reach of the planned  charged lepton flavor violating experiments.
Interesting lower bounds  are predicted that $B(Z\rightarrow \overline{l} l')\gtrsim  10^{-16}-10^{-14}(10^{-14})\times(1\tev\cdot m_S/7 m_\Delta^2)^2$ and $B(l\rightarrow l' \gamma)\gtrsim 10^{-17}-10^{-16}(10^{-18}-10^{-16})\times(1\tev\cdot m_S/7 m_\Delta^2)^2$ for neutrino masses being the normal(inverted) hierarchical  pattern.
The type of neutrino mass hierarchy could also be determined by measuring the charged lepton flavor violating double ratios.
Moreover,  definite leptoquark decay branching ratios are predicted  when there is no Yukawa interaction between the right-handed fermions and $\Delta$ ( the  branching fraction of $\Delta$ to a charged lepton and a quark is 50\%), which could help refine the collider search limit on the scalar leptoquark mass.
}


\begin{document}

\maketitle
\flushbottom

\section{Introduction}
It is now well established that at least two of the active neutrinos are massive.
New physics beyond the Standard Model(SM) is required to give
small but nonzero neutrino masses. A straightforward remedy is adding the right-handed neutrino(s) to the SM so that the active neutrinos can acquire Dirac masses after the SM electroweak symmetry breaking as what other charged fermions do. However,  additional mass suppression mechanisms or  very tiny Yukawa couplings, $\lesssim 10^{-12}\times Y_{top}$, are required  to bring down the resulting Dirac neutrino masses to the  sub-eV level.   Alternatively, Majorana neutrino masses are sought to alleviate the problem of huge hierarchy among the Yukawa couplings in the Dirac neutrino cases.
  Whatever the UV origin of Majorana neutrino mass is, the key is to  generate
the dimension-5 Weinberg effective operator\cite{Weinberg:1979sa}, $(LH)^2$, where $L$ and $H$ are the SM lepton doublet and the Higgs doublet, respectively, at the low energy.
 The Weinberg operator conserves the baryon number but violates lepton number by two units.
Since all the SM interactions at the low energies conserve both baryon and lepton numbers,
the new interactions responsible for generating the Majorana neutrino masses  must break the lepton number and the relevant new degree(s) of freedom must carry lepton number.
If the relevant new fields also carry nonzero baryon number, there are no tree-level contributions to the Weinberg operator leading  to a nature loop suppression to bring down the resulting Majorana neutrino masses.
Therefore, leptoquark, a boson which carries both lepton number and baryon number, is one of the well-motivated candidates to generate small Majorana neutrino masses without excessive fine tuning.
Moreover, since leptoquark participates strong interaction, it would be interesting that the new particles relevant to  the neutrino mass generation mechanism could be directly  probed at the hadron colliders.
However,   it is impossible to generate the desired Weinberg operator by  using only one leptoquark because the new interaction vertices always come in conjugated pairs.
Something else in the loop(s) which carries baryon number  must also be utilized  to have zero net baryon number and non-vanishing lepton number at the end.
The  di-quark, a boson which carries $2/3$ of baryon number, is one of the candidates to work with leptoquark
for generating the Weinberg operator\footnote{It is also perfectly possible to generate nonzero neutrino masses with two leptoquarks with different lepton and  baryon numbers, see for example\cite{LQ_1_loop_Wise, LQ_1_loop_2}. }.
Neutrino masses aside, the leptoquark and di-quark are  common in many new physics models where the lepton number or the baryon number is not conserved\cite{LQ:Buchmuller,LQ:Dimopoulos,LQ:Dimopoulos2,LQ:Eichten,LQ:Angelopoulos,DQ:Hewett}, such as the grand-unified theories, technicolor and composite models.
Yet without positive results,  leptoquark and di-quark had been eagerly searched for since the  1980's.
At the colliders, the leptoquark and di-quark could be produced and studied directly. However, the  decay rates strongly depend on the unknown couplings between the leptoquark/di-quark and the SM fermions. Thus, the bounds are usually given with specific assumptions on their couplings to the SM fermions, see \cite{ATLAS-1st,ATLAS-2nd,ATLAS-3rd,CMS-1st,CMS-2nd,CMS-3rd,HERA-1st,LQ-CMS-ATLAS,DQ-mass}.
On the other hand, flavor changing processes could be mediated by the leptoquark or di-quark at the tree-level. Strong constraint
can be indirectly derived from the low energy flavor changing experiments\cite{2L2Q}.

Recently, an interesting application of utilizing the scalar leptoquark and scalar di-quark  to generate the neutrino masses was discussed by \cite{Masaya}.
 In \cite{Masaya},   one scalar leptoquark, $\Delta$ with SM  $SU(3)_c\times SU(2)_L\times U(1)_Y$ quantum number $(3,1,-1/3)$, and one scalar di-quark, $S$ with SM quantum number $(6,1,-2/3)$,  were augmented to the SM particle content  and the neutrino masses can be generated through the two-loop radiative corrections\footnote{See \cite{Queiroz:2014pra} for a recent discussion on the   potential connection between $\Delta$ and the dark matter. }.
This two-loop mechanism is very similar to that in the Zee-Babu model\cite{Zee-Babu} except
that $S/\Delta$ replaces the role of the doubly/singly  charged scalar in Zee-Babu model.  From now on this model is referred as the colored Zee-Babu Model(cZBM). In the cZBM, the resulting neutrino mass matrix pattern and the mixing angles are determined by  $Y_L$ and $Y_S$, the Yukawa couplings between leptoquark and di-quark and the SM fermions, see Eq.(\ref{eq:Yukawa}).
Again, $Y_S$ and $Y_L$ are arbitrary and a priori unknown. To proceed, we consider the case that the symmetric $Y_S$'s are democratic and the magnitudes of the six $(Y_S)_{ij}$, where $i,j=1,2,3$ are the flavor indices,  are of the same order. This could be  realized in the extra-dimensional models with the right-handed down-type quark bulk wave functions cluster together in the extra spatial dimension(s), for applying the geometric setup to generate a special 4-dimensional Yukawa pattern see for example  \cite{Chang:RS,Chang:SF}. With this working assumption and the fact that $m_b\gg m_s \gg m_d$, the $Y_L$ can be determined with some reasonable requirements which will be discussed later.
To accommodate all the neutrino data, the tree-level flavor violating processes will be inevitably mediated by  $\Delta$ with the realistic $Y_L$ Yukawa couplings.
Moreover, the rates of these resulting tree-level and also those flavor violating processes induced at the loop level must comply with the current experimental bounds.
 In addition to $Y_L$, $\Delta$ also admits  Yukawa couplings, $Y_R$, which couple $\Delta$ to the right-handed leptons and quarks, see Eq.(\ref{eq:Yukawa}).
Since both $Y_L$ and $Y_R$ contribute to the tree-level flavor violating processes incoherently,  $Y_R=0$ is assumed  to minimize those rates.
A comprehensive numerical study is performed to search for the realistic configurations.
 We find that sizable portion of the realistic solutions overlap with  the designed sensitivities of the forthcoming lepton flavor violation experiments.
Moreover, for  the neutrino masses in both the normal hierarchy and inverted hierarchy, there are  interesting and definite lower bounds on $B(Z\rightarrow\overline{ l} l')$ and $B(l\rightarrow l' \gamma)$  which could be falsified in the future.
Also, the type of neutrino mass hierarchy can be determined if the charged lepton flavor violating double ratios are measured to be within some specific ranges.
If $Y_R=0$,  the model has concrete predictions  for the scalar leptoquark decay branching ratios for both  neutrino mass hierarchies. This will help refine the collider search limit on the scalar leptoquark  mass for the $\beta=1/2$ case.

The paper is organized as follows. A more detailed discussion on the model is given in section 2.
In section 3, we study the connection between the neutrino masses and $Y_L$, and the tree-level flavor violating processes as well.  The loop-induced flavor violating processes are discussed in section 5.   The numerical study are dealt with and discussed in section 5. Finally, the conclusions are summarized in section 6.

\section{Model and neutrino mass}

\begin{figure*}[h]
\begin{center}
\includegraphics[width=8cm]{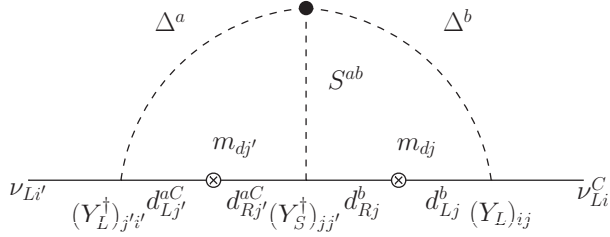}
\caption{The 2-loop neutrino mass generated from colored scalar. Where $a,b,c$ are the $SU(3)$ indices.} \label{neutrino-mass}
\end{center}
\end{figure*}

As mentioned in the previous section, the SM is extended by adding  $S$ and $\Delta$.
After rotating the lepton fields into their  weak basis and the  quarks into their mass basis, the most general gauge invariant Yukawa interaction associated with $S$ and $\Delta$ is
\begin{equation}
\mathscr{L}_Y=-\left[\overline{L_{i}^C}(Y_L)_{ij}i\sigma_2 Q_j
+\overline{(\ell_{Ri})^C}(Y_R)_{ij}u_{R j}\right] \Delta^{*}
-\overline{(d_{Ri})^{C}} (Y_s)_{ij} d_{Rj} S^{*}
+ y^\Delta_{ij}  \overline{(u_{R i})^C} d_{R j} \Delta
+h.c.\label{eq:Yukawa}
\end{equation}
where $i,j$ are the flavor indices and the $SU(3)$ indices are suppressed.
Apparently $Y_S$ is symmetric in the flavor space while there is no such constraints on $Y_L$,  $Y_R$, and $y^\Delta$.
Moreover, the lagrangian admits a gauge invariant triple coupling term: $ (\mu \Delta^* \Delta^* S +h.c.)$.
As shown in Fig. \ref{neutrino-mass}, the neutrino masses will receive nonzero contributions through 2-loop quantum corrections if both $Y_L$ and $Y_S$ present. If one writes the effective Lagrangian for neutrino masses as $-\frac12 \overline{\nu_{Li}^C}(M_\nu)_{ij} \nu_{Lj}$, the neutrino mass matrix can be calculated to be
\begin{align}
\label{eq:nu_mass_elements}
&(M_\nu)_{ii'}= 24\mu (Y_L)_{ij}m_{dj} I_{jj'} (Y_s^\dagger)_{jj'}m_{dj'}(Y^T_{L})_{j'i'}\,,\\
&I_{jj'}=\int \frac{d^4k_1}{(2\pi)^4}\frac{d^4k_2}{(2\pi)^4}\frac{1}
{(k_1^2-m_{dj}^2)}\frac{1}
{(k_1^2-m_{\Delta}^2)}\frac{1}
{(k_2^2-m_{dj'}^2)}\frac{1}
{(k_2^2-m_{\Delta}^2)}\frac{1}{(k_1+k_2)^2-m_S^2}\,.
\end{align}
Note that the two-loop integral is similar to the one in the Zee-Babu model\cite{ZBM,Analytical-ZBM}.
When the down-type quark mass is much lighter than colored scalars, the integral is flavor independent and it can be simplified to
\begin{align}
&I_{jj'}\simeq I_\nu \equiv
\frac{1}{(4\pi)^4}\frac{1}{M^2}\frac{\pi^2}{3}\tilde{I}\left(\frac{m_S^2}{m_\Delta^2}\right),\qquad
M\equiv \mathrm{max}(m_\Delta, m_S)\,,\\
&\tilde{I}(x)=\left\{\begin{array}{ll}
1+\frac{3}{\pi^2}(\ln^2 x-1) & \mathrm{for}~ x\gg 1\,,\\
1& \mathrm{for}~ x\to 0\,.
\end{array}\right.
\end{align}
For the later use, it is convenient to write the neutrino mass matrix in a compact form
\begin{equation}\label{eq_seesaw_relation}
M_{\nu}=Y_L\omega Y_L^T,
\end{equation}
with the matrix $\omega_{jj'}\equiv 24\mu I_{\nu}m_{j}m_{j'}(Y_s^\dagger)_{jj'}$. Qualitatively speaking, the resulting
neutrino mass is about
\beq
m_\nu \sim {\mu m_b^2 Y_L^2 Y_S \over 32 \pi^2 M^2} \sim 0.06 \mbox{eV} \times\left( {Y_L^2 Y_S \over 10^{-6} }\right)
\times\left( {  \mbox{TeV}\over M^2/ \mu }\right)\,.
\label{eq:BOE_nu_mass}
\eeq
One sees that, due to the 2-loop suppression, with a typical values $Y_L,Y_S \sim 0.01$ and $\mu, M \sim 1$ TeV, the sub-eV neutrino mass can be easily achieved without excessively fine tuning.

However, the simultaneous presence of $Y_{L/R}$ and $y^\Delta$ leads to tree-level proton decay as pointed out in \cite{LQ-Wise}.
A very small $y^\Delta_{11}$ is enough to avoid the rapid proton decay problem. Alternatively, the $y^\Delta$ term can be
eliminated by imposing some ad hoc symmetry. For example, this term can be turned off without upsetting all other interactions if  some $Z_2$ parities $\{-,-,+, +, +, -,+\}$  are assigned to $\{ L,  l_R, Q, u_R, d_R, \Delta, S \}$, respectively.
Hence, we leave the proton decay problem aside and simply set $y^\Delta=0$ in this study.

The most general renormalizable scalar potential including $S$ and $\Delta$ is
\begin{eqnarray}
\mathscr{V}&=&
- \mu_H^2 (H^\dagger H)
 +m_\Delta^{2}\Delta^\dagger \Delta +
m_S^{2} \mbox{Tr} S^\dagger S
+\lambda(H^\dagger H)^2 + \lambda_\Delta (\Delta^\dagger \Delta)^2
 +\lambda_S (\mbox{Tr} S^\dagger S)^2
 \label{potential}
\\
 &&  +\lambda_1 (\Delta^\dagger \Delta)(H^\dagger H)
+\lambda_2 \mbox{Tr}(S^\dagger S)(H^\dagger H)
+\lambda_3 \mbox{Tr}(S^\dagger S)(\Delta^\dagger \Delta)
+ \left(\mu \Delta^{*}
\Delta^{*}S + h.c.\right)\nonumber
\end{eqnarray}
where the trace is over the color indices.
The details of the scalar potential are not  relevant for the later discussion.  We note by passing that only the SM Higgs doublet can acquire a nonzero vacuum expectation value, $\langle H \rangle =v/\sqrt{2} $,  and being solely responsible for the electroweak symmetry breaking(EWSB). The tree-level masses of $\Delta$ and $S$ are shifted after EWSB with $ m_\Delta^{2} \rightarrow m_\Delta^{2} + \lambda_1 v^2/2$ and $ m_S^{2} \rightarrow m_S^{2} + \lambda_2 v^2/2$.
To proceed, we need  $m_\Delta$ and $m_S$ after EWSB as input.
Since $S$ and $\Delta$ participate strong interactions, they are best searched for at the hadron colliders but so far none has been found yet.
Depending on their couplings to the SM fields, some lower bounds on $m_S$ and $m_\Delta$ were obtained from the null result of collider searches. The current lower bounds on $m_\Delta$ are summarized in   Table.\ref{tab:LQ-mass}.
\begin{table*}[h]
\caption{
Summary of  leptoquark mass lower bound (in unit of $\mathrm{GeV}$) from direct search with $95\%$ CL.
The values in parentheses are for $\beta=0.5$, and  $\beta=1$ otherwise. The leptoquark decays branching ratios into $lq$ and $\nu q$ are denoted as $\beta$ and $(1-\beta)$, respectively,
and $\lambda$ is the Yukawa coupling for $l q \Delta$. The leptoquark is assumed to decay into leptons within only one specific generation.
}
\begin{center}
\begin{tabular}{c|c|c|c}
\hline\hline
 & First generation  & Second generation & Third generation \\
 \hline
%\hline
 CMS & $1005(845)$ \cite{CMS-1st} & $1070 (785)$ \cite{CMS-2nd}   & $634$  \cite{CMS-3rd} \\
 ATLAS  & $ 660 (607)$ \cite{ATLAS-1st} &  $685(594)$ \cite{ATLAS-2nd}    &  $534$ \cite{ATLAS-3rd} \\
 ZEUS & $699 (\lambda=0.3)$\cite{HERA-1st} & & \\
\hline\hline
\end{tabular}
\end{center}
\label{tab:LQ-mass}
\end{table*}
For an $E_6$-type diquark, CMS study gives $m_{S}>6\mathrm{TeV}$ \cite{DQ-mass}.
These limits are very sensitive to the assumptions of decay branching fraction  as well as the flavor dependant coupling strengthes.
Hence, in the following numerically analysis, we take $m_S=7\,\mathrm{TeV}$ and
 $m_\Delta=1\,\mathrm{TeV}$ as the benchmark values\footnote{
Since $S$ and $\Delta$ are also charged under SM $SU(3)$ and $U(1)_Y$, their 1-loop contributions alter the SM $h VV'$ couplings where $VV'=\{\gamma\gamma, \gamma Z, gg\}$.
Following the analysis in \cite{Chang-Ng-Wu,CPX,CGHT} also the data form\cite{ATLAS-hdiphoton,ATLAS-hgammaZ,CMS-hdiphoton,CMS-hgammaZ}, we find that the corrections to the signal strengths
are not significant, $0.96<\mu_{\gamma\gamma, \gamma Z}<1.2$, for $ m_\Delta\in[1,3]$ TeV and $m_S\in[6,8]$ TeV.}.

The triple $\Delta\Delta S$ coupling  generates 1-loop correction $\sim \frac{\mu^2}{16\pi^2}\log(\mu^2/m_X^2)$ to $m_X^2$ where $X=S, \Delta$. For these quantum corrections to be perturbative, one needs roughly  $ |\mu ^2 \log(\mu^2/m_X^2)|\leq 16\pi^2 m_X^2$
\footnote{ From the Eq.(\ref{eq:BOE_nu_mass}), $\mu$ also has a weak lower bound $|\mu| \gtrsim 10^{-6}\mbox{TeV}\times (M/\mbox{TeV})^2 $ if $Y_L$ and $Y_S$ are required to be less than $1.0$. }.
On the other hand, the dimensionful parameters in the same scalar potential are expected to be around the same order.
 These considerations led to similar estimations and  $\mu=(0.1- 1)\,\tev$ is assumed in this study.

At the tree-level, the decay channels for leptoquark  are $\Delta \rightarrow \ell_i u_j$ and $\Delta \rightarrow \nu_i d_j$. For di-quark, it decays into
 $d_i d_j$, and $\Delta \Delta$ if kinematically allowed.
Given that $m_S, m_\Delta \gg m_t$, all the SM final states can be treated massless and the decay widthes
can be calculated to be
\begin{eqnarray}
\Gamma_{\Delta} &=& \sum_{i,j}
					\left[ \Gamma(\Delta \rightarrow \ell_i u_j) + \Gamma(\Delta \rightarrow \nu_i d_j) \right]
			    \sim  \frac{ m_{\Delta}}{16 \pi} \sum_{i,j}					
				 \left( 2 \vert (Y_L)_{ij} \vert^2 + \vert (Y_R)_{ij} \vert^2  \right)\,,\\
\Gamma_{S}   &=&  \Gamma(S \rightarrow \Delta \Delta)\times \theta(m_S-2 m_\Delta) + \sum_{ij } \Gamma(S \rightarrow d_i d_j)  \nonumber \\
		     &\sim &  \frac{m_S}{8 \pi}  \left\lbrace
				  \left( 1-4 \frac{m^2_{\Delta}}{m_S^2} \right)^{1/2} \left( \frac{\mu}{m_S}\right)^2\times \theta(m_S-2 m_\Delta)
				 + \sum_{i,j} \vert (Y_S)_{ij} \vert^2 	  \right\rbrace.
\end{eqnarray}

\section{Neutrino Masses and the Tree-level Flavor Violation}
As discussed before, it is assumed that there is no hierarchy among the $Y_S$'s.
Since $m_b \gg m_s \gg m_d$, the matrix $\omega$ can be broken into the leading and sub-leading parts and $\omega =\omega^{(0)}+\omega^{(1)}$, where
\beq
\omega^{(0)}= 24 \mu I_\nu \times   \left( \begin{array}{ccc}
                                              0 & 0 & 0 \\
                                              0 & 0 & m_b m_s (Y_S)_{23}^*\\
                                              0 & m_b m_s (Y_S)_{23}^* & m_b^2 (Y_S)_{33}^*
                                            \end{array}
\right)\,,
\eeq
and ${\cal O} \left(\frac{\omega^{(1)}}{\omega^{(0)} }\right) \sim  {\cal O}\left( \frac{m_d}{m_b} \right)$.
It is easy to check that the leading neutrino mass matrix $M_\nu^{(0)} = Y_L \omega^{(0)} Y_L^{T}$ is of rank-2 and $\det M_\nu^{(0)}=0$. Hence,  at least one of the active neutrinos is nearly massless, $\sim (m_d/m_b)\times \mbox{max}(m_\nu)$, and the scenario of  quasi-degenerate neutrinos is disfavored in the cZBM.
At leading order, $(Y_L)_{11,21,31}$ do not enter $M_\nu^{(0)}$ at all.
Therefor, for either normal hierarchy (NH) or the inverted hierarchy(IH) type of the neutrino masses, the eigenmasses are
\begin{itemize}
\item   NH:
\begin{equation}
m_1\simeq 0\,,   m_2 \simeq \sqrt{\Delta m_{21}^2}\,,   m_3 \simeq  \sqrt{\Delta m^2 + \frac{\Delta m_{21}^2}{2} }\,,
\end{equation}
and
\item  IH:
\begin{equation}
m_3\simeq 0\,,  m_1 \simeq  \sqrt{|\Delta m^2| - \frac{\Delta m_{21}^2}{2} }\,,  m_2 \simeq  \sqrt{|\Delta m^2| + \frac{\Delta m_{21}^2}{2} }\,,
\end{equation}
\end{itemize}
where  $\Delta m^2 \equiv m_3^2 - ( m_1^2 + m_2^2 )/2$.
Moreover, the absolute values of neutrino mass can be obtained by plugging in the well determined neutrino data\cite{PDG} listed  in Table \ref{tab:nu_exp}.
For  NH, $ m_2 \sim 0.00868$ eV and $m_3\sim 0.0496$ eV, and  for IH, $m_1\sim 0.0483$ eV and $m_2\sim 0.0492$ eV. For both cases, the total sum of neutrino masses automatically agrees with the  limit that $\sum m_{\nu}<0.23$ eV at $95\%$ C.L.
from the cosmological observation\cite{nu-mass}.
\begin{table*}[h]
\caption{
The global-fit neutrino data with $1\sigma$ deviation\cite{PDG}. }
\begin{center}
\begin{tabular}{c|c}
\hline\hline

$\sin^2 \theta_{23}$
						&	$0.437^{+0.033}_{-0.023}$ \quad (NH)  \\
						&	 $0.455^{+0.039}_{-0.031}$ \quad (IH)  \\							

$\sin^2 \theta_{13}$  	&  $0.0234^{+0.0020}_{-0.0019}$\quad (NH) \\
					& $0.0240^{+0.0019}_{-0.0022}$ \quad (IH)  \\
$\sin^2 \theta_{12}$  &  $0.308^{+0.017}_{-0.017}$ \\
$\Delta m_{21}^2$  & $ \left(7.54^{+0.26}_{-0.22}\right) \times 10^{-5} \mathrm{eV}^2$  \\

$|\Delta m^2|$

					& $\left(2.43^{+0.06}_{-0.06}\right)\times 10^{-3}\mathrm{eV}^2$ \quad (NH) \\
					& $\left(2.38^{+0.06}_{-0.06}\right)\times 10^{-3}\mathrm{eV}^2 $ \quad (IH) \\
$\delta/\pi$      & 	$1.39^{+0.38}_{-0.27}$ \quad (NH) \\
                 &	$1.31^{+0.29}_{-0.33}$ \quad (IH) \\
						\hline\hline
\end{tabular}
\end{center}
\label{tab:nu_exp}
\end{table*}

Once $m_{1,2,3}$ are fixed, the neutrino mass matrix can be worked out reversely by
\beq
\label{eq:mnu_diag}
M_\nu = U_{PMNS}^*   \left( \begin{array}{ccc}
                                              m_1 & 0 & 0 \\
                                              0 & m_2 &0\\
                                              0 & 0& m_3
                                            \end{array}
\right) U_{PMNS}^\dag\,.
\eeq
The standard parametrization is adopted that
\beq
\label{eq:definition_of_pmns}
 U_{\mathrm{PMNS}}=\left(\begin{array}{ccc} 1 & 0 & 0\\
0 & c_{23} & s_{23}\\
0 & -s_{23} & c_{23}\end{array}\right)\left(\begin{array}{ccc}
c_{13} & 0 & s_{13} e^{-i\delta} \\
0 & 1 & 0 \\
-s_{13} e^{i\delta} & 0 & c_{13}\end{array}
\right)\left(\begin{array}{ccc}
c_{12} & s_{12} & 0 \\
-s_{12} & c_{12} & 0 \\
0 & 0 & 1\end{array}\right)
\begin{pmatrix}
1 & 0 & 0
\\
0 & e^{  i \alpha_{21}/2} & 0
\\
0 & 0 & e^{  i \alpha_{31}/2}
\end{pmatrix}
\eeq
where $c_{ij}$ and $s_{ij}$ represent $\cos\theta_{ij}$ and $\sin\theta_{ij}$, respectively.
In the case of Majorana neutrinos, $\alpha_{21}$ and $\alpha_{31}$ are the extra CP phases that cannot be determined from the oscillation experiments.
For simplicity, all $Y_S$'s are assumed to be real and the 2 Majorana CP phases  will not be discussed in this paper.
The leading order neutrino mass matrix has 5(=6-1) independent entries\footnote{ The symmetric neutrino matrix $M_\nu^{(0)}$ has 6 elements minus 1 constraint that its determinant is zero.  }.
With the democratic $Y_S$ assumption, the effective Majorana mass for $(\beta\beta)_{0\nu}$-decay $m_{ee} \sim 0.0018$ eV for the NH case. For the IH case,  $m_{ee} \sim 0.0479$ eV which is
 within the sensitivity of the planned $(0\nu\beta\beta)$ detectors with  $\sim 1$ ton of isotope\cite{Bilenky:2012qi}.
Furthermore, the lightest neutrino mass is $\sim {\cal O}(10^{-5} \mbox{eV})$ for both IH and NH cases.  For a given set of parameters, $\{\mu, m_S, m_\Delta, (Y_S)^{(0)}_{23}, (Y_S)^{(0)}_{33}, (Y_L)_{13} \}$, all the other 5 complex Yukawa couplings $(Y_L)_{ij} (j\neq 1)$ can be completely determined up to two signs by the leading $M_\nu^{(0)}$.
For a real $(Y_L)_{13}$, one has
\beqa
\label{eq:YL_leading}
(Y_L)_{23}^{(0)} &=& \frac{(Y_L)_{13}}{(M_\nu)_{11}}\left[(M_\nu)_{12}\pm \sqrt{(M_\nu)_{12}^2-(M_\nu)_{11}(M_\nu)_{22}}\right]\,,\nonr\\
(Y_L)_{33}^{(0)} &=& \frac{(Y_L)_{13}}{(M_\nu)_{11}}\left[(M_\nu)_{13}\pm \sqrt{(M_\nu)_{13}^2-(M_\nu)_{11}(M_\nu)_{33}}\right]\,,\nonr\\
(Y_L)_{12}^{(0)} &=& { (M_\nu)_{11} -B_\nu m_b^2 (Y_L)_{13}^2  (Y_S)_{33}^{(0)}  \over 2 B_\nu m_b m_s (Y_L)_{13} (Y_S)_{23}^{(0)} }\,,\nonr\\
(Y_L)_{22}^{(0)} &=& { (M_\nu)_{22} -B_\nu m_b^2 [(Y_L)_{23}^{(0)}]^2  (Y_S)_{33}^{(0)}  \over 2 B_\nu m_b m_s (Y_L)_{23}^{(0)} (Y_S)_{23}^{(0)} }\,,\nonr\\
(Y_L)_{32}^{(0)} &=& { (M_\nu)_{33} -B_\nu m_b^2 [(Y_L)_{33}^{(0)}]^2  (Y_S)_{33}^{(0)}  \over 2 B_\nu m_b m_s (Y_L)_{33}^{(0)} (Y_S)_{23}^{(0)} }\,,
\eeqa
where $B_\nu=24 \mu I_\nu$. Again, $(Y_L)_{11,21,31}$ do not enter $M_\nu^{(0)}$ at all; they are arbitrary at this level and will be determined in the next order perturbation.
This approximation largely saves the work of numerical study and lays out the base for higher order perturbations beyond $\omega^{(0)}$.

The most important next to leading contribution to $M_\nu$ comes from $(Y_S)_{13}$. If one also perturbs $(Y_S)_{23,33}$ around
$(Y_S)^{(0)}_{23,33}$  with $(Y_S)_{23,33}^{(1)}=(Y_S)^{(0)}_{23,33}+\delta_{23,33}$, the  consistent solution to $(Y_L)_{i1}$ for a given set $\{(Y_S)_{13}, \delta_{23},\delta_{33}\}$ are:
\beqa
\label{eq:YL_next}
(Y_L)_{11}^{(1)} &=& -\frac{m_s}{m_d}\frac{\delta_{23}}{(Y_S)_{13}}(Y_L)_{12}^{(0)}-\frac{m_b}{2m_d}\frac{\delta_{33}}{(Y_S)_{13}}(Y_L)_{13}\,,\nonr\\
(Y_L)_{21}^{(1)} &=& -\frac{m_s}{m_d}\frac{\delta_{23}}{(Y_S)_{13}}(Y_L)_{22}^{(0)}-\frac{m_b}{2m_d}\frac{\delta_{33}}{(Y_S)_{13}}(Y_L)_{23}^{(0)}\,,\nonr\\
(Y_L)_{31}^{(1)} &=& -\frac{m_s}{m_d}\frac{\delta_{23}}{(Y_S)_{13}}(Y_L)_{32}^{(0)}-\frac{m_b}{2m_d}\frac{\delta_{33}}{(Y_S)_{13}}(Y_L)_{33}^{(0)}\,.
\eeqa

 With only a handful of free parameters, all the leptoquark left-handed Yukawa can be reasonably determined solely by the neutrino data.
However, further checks are needed to determine  whether the above solution is  phenomenologically viable.
Next, the tree-level flavor violation will be discussed.

\begin{figure*}[h]
\begin{center}
\includegraphics[width=8cm]{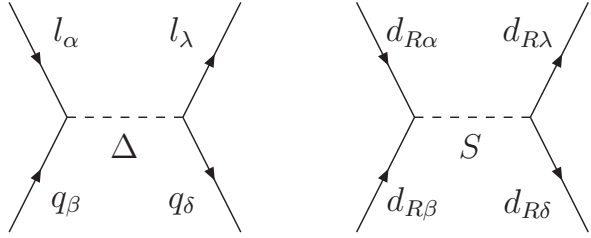}
\caption{Tree-level flavor violation mediated by leptoquark and diquark. }
\label{fig:TL_FV}
\end{center}
\end{figure*}

The leptonic and quark flavor violating processes will be generated by exchanging $S$ and $\Delta$ at the tree-level, see Fig.\ref{fig:TL_FV}.
Since $S,\Delta$ are heavy, they can be integrated out below the EWSB scale. After Fierz transformation, we obtain
\begin{eqnarray}
\begin{aligned}
\triangle \mathcal{L}_{\mbox{eff}} =& \left[\frac{  (Y_{L}^{*})_{ml} (Y_{R})_{ij} }{2 m_{\Delta}^2}
							\left(
							-\overline{\nu_{m}} \PP_{R} \ell_{i} \cdot \overline{d_{l}^{a}} \PP_{R} u_{j}^{a}
							 +\overline{\ell_{m}} \PP_{R} \ell_{i} \cdot \overline{u_{l}^{a}} \PP_{R} u_{j}^{a}
							 \right)
							 +h.c. \right] \\
				  & -\left[\frac{ (Y_{L}^{*})_{ml}  (Y_{L})_{ij} }{2 m_{\Delta}^2} \overline{\nu_{m}} \gamma^{\mu} \PP_{L} \ell_{i} \cdot \overline{d_{l}^{a}} \gamma_{\mu} \PP_{L} u_{j}^{a}
				     + h.c. \right]
				      \\
				  &  +	\frac{ (Y_{L}^{*})_{ml} (Y_{L})_{ij}  }{2 m_{\Delta}^2}
				  		\left(
						\overline{\nu_{m}} \gamma^{\mu} \PP_{L} \nu_{i} \cdot \overline{d_{l}^{a}} \gamma_{\mu} \PP_{L} d_{j}^{a}				
						+  \overline{\ell_{m}} \gamma^{\mu} \PP_{L} \ell_{i} \cdot \overline{u_{l}^{a}} \gamma_{\mu} \PP_{L} u_{j}^{a}
					 	\right) \\
				  &   +\frac{ (Y_{R}^{*})_{ml} (Y_{R})_{ij}  }{2 m_{\Delta}^2} \overline{\ell_{m}} \gamma^{\mu} \PP_{R} \ell_{i} \cdot \overline{u_{l}^{a}} \gamma_{\mu} \PP_{R} u_{j}^{a}
				     \\
			& +
			\left[			
			\frac{ (Y_{L}^{*})_{ml} (Y_{R})_{ij}  }{8 m_{\Delta}^2}
			\left(			
			\overline{\nu_{Lm}} \sigma^{\mu\nu} \PP_{R} \ell_{i} \cdot \overline{d_{l}^{a}} \sigma_{\mu\nu} \PP_{R}  u_{j}^{a}
			-\overline{\ell_{Lm}} \sigma^{\mu\nu} \PP_{R} \ell_{i} \cdot \overline{u_{l}^{a}} \sigma_{\mu\nu} \PP_{R}  u_{j}^{a}
			\right) + h.c.\right]\\
& + \frac{(Y_s)_{ij}(Y_s^\dagger)_{lm} }{2m_S^2}\left[\overline{d_m^a}\gamma^\mu\PR d_i^a\right]
\left[\overline{d_l^b}\gamma_\mu\PR d_j^b\right]
\label{eq:TLFV}
\end{aligned}
\end{eqnarray}
where $a,b$ are the color indices.

There are way too many new free parameters and rich phenomenology in the most general model.
To simplify the discussion and to extract the essential physics, we consider the case that the new physics has minimal tree-level flavor violation(TLFV). Note that the TLFV contributions  from different chiral structures always add incoherently.   To minimize the total TLFV we need to suppress the TLFV from each chiral structure as much as possible.  Let's concentrate on the
purely left-handed operators first. Observe that  (1) A trivial flavor violation free  solution  is that with $(Y_{L})_{ij}(Y_{L}^\dagger)_{lm} \propto \delta_{im} \delta_{j l}$. It is obvious that these kind of solutions allow only one nonzero entry of $Y_{L}$, as can be easily
seen by looking at Fig.\ref{fig:TL_FV}(a). It always leads to 2 massless neutrinos which has been excluded by the current neutrino oscillation data.
(2) If the requirement is relaxed to  $(Y_{L})_{ij}(Y_{L}^\dagger)_{lm} \propto \delta_{im}$ (no leptonic TLFV) or $ (Y_{L})_{ij}(Y_{L}^\dagger)_{lm} \propto \delta_{j l}$ ( no quark TLFV), only one  row or one  column of $Y_{L}$ can be nonzero\footnote{Neutrino masses aside, similar conclusions also apply to the $Y_R$ matrix for the purely right-handed operators.}
and the resulting neutrino masses have two zeros again.  However, only $Y_L$'s are relevant to the neutrino masses.
One can set $Y_R =0$ to minimize  the TLFV, and use the 9 remaining  $Y_L$'s to accommodate the neutrino masses. Then the lower bound on each flavor violation process can be found since any nonzero $Y_R$ will add to it.

It is very easy to build a  model with $Y_R\ll 1$ or $Y_R=0$ and we supplement with two examples.
Example one is to introduce an extra $U(1)_x$  with two SM like Higgs doublets, $H_1$ and $H_2$.
Then the $U(1)_x$ charge assignment  $\{\alpha_1,\alpha_2,\alpha_2,\alpha_3,\alpha_4 ,\alpha_5, \alpha_6, \alpha_7 \}$ for  $\{ Q, u_R, d_R, L, e_R, H_1, H_2, \Delta\}$ with $ \alpha_5= \alpha_1-\alpha_2$, $\alpha_6=\alpha_3-\alpha_4$, $\alpha_7=-(\alpha_1+\alpha_3)$, $\alpha_7 \neq 2 \alpha_2$ and $\alpha_7 \neq (\alpha_2+\alpha_4)$ will kill $Y_R$ (and also $y^\Delta$) but still allow the charged fermions to acquire the Dirac masses from their Yukawa couplings with $H_1$ or $H_2$. There are other issues needed to be considered in this setup. For example whether the $U(1)_x$ is global or local and whether it is free of anomaly. But these issues do not concern us since they are not relevant to this study and there are well-known model-building machineries available to deal with these problems.
The second example is promoting the 4-dimensional model into a higher-dimensional version. If the wave functions of $l_R$ and $u_R$ in the extra spacial dimension(s) are well separated, like in \cite{Chang:SF}, or have very little overlapping, like in\cite{Chang:RS}, the resulting $Y_R$ is negligible. Anyway, here  $Y_R=0$ is taken as a phenomenology assumption which minimizes the TLFV. Some remarks on the case of $Y_R\neq 0$ will be discussed in next section.

A model independent analysis of the effective four-fermion operators was done by\cite{2L2Q}.
The $90\%$ C.L. upper limits on the normalized Wilson coefficient $\epsilon_{ijkl}$ (it is not the totally anti-symmetric tensor),
\begin{equation}
\epsilon_{ijkn} \equiv \frac{ (Y_L)_{ik} (Y_L)_{jn}}{4 \sqrt{2} G_F m_\Delta^2}\,,
\end{equation}
 for each 4-fermi operator are extracted and listed in Table \ref{tab:2L2Q-YL}.
We have
\begin{equation}
(Y_L)_{ik} (Y_L)_{jn} < 4 \sqrt{2} G_F m_\Delta^2 \epsilon_{ijkn}^\text{max} \sim 65.98 \times \epsilon_{ijkn}^\text{max} \times\left({ m_\Delta \over 1 \mbox{TeV}}\right)^2\,.
\end{equation}
\begin{table*}[t]
\caption{
 The 90\%C.L. upper limits on $\epsilon_{ijkn}$ from \cite{2L2Q}. Here, the dimensionless quantities  $\epsilon_{ijkn} \equiv (Y_L)_{ik} (Y_L)_{jn} / ( 4 \sqrt{2} G_F m_\Delta^2 )$, where  $i,j$($k,n$)are lepton(quark) flavor indices.
%The limit $\epsilon_{e\mu 33}$ has been updated with new experimental data.
}
\begin{center}
\begin{tabular}{|c|c||c|c||c|c|}
\hline
$\epsilon_{ e e 1 1 }  $ & $ 10^{-3}$
& $ \epsilon_{ e e 1 2 }  $ & $ 9.4 \times 10^{-6} $ & $ \epsilon_{ e e 1 3 } $ & $ 3.9 \times 10^{-3} $ \\
\hline
$\epsilon_{ e e 2 2 } $ & $ 10^{-2} $ & $ \epsilon_{ e e 2 3 } $ & $ 10^{-3} $ & $ \epsilon_{ e e 3 3 } $ & $ 9.2 \times 10^{-2} $ \\
\hline
\hline
$\epsilon_{ \mu \mu 1 1 } $ & $ 7.3 \times 10^{-3} $ & $ \epsilon_{ \mu \mu 1 2 }  $ & $ 9.4 \times 10^{-6} $ & $ \epsilon_{ \mu \mu 1 3 } $ & $ 3.9 \times 10^{-3} $ \\
\hline
$\epsilon_{ \mu \mu 2 2 } $ & $ 1.2 \times 10^{-1} $ & $ \epsilon_{ \mu \mu 2 3 } $ & $ 10^{-3} $ & $ \epsilon_{ \mu \mu 3 3 } $ & $ 6.1 \times 10^{-2} $ \\
\hline
\hline
$ \epsilon_{ \tau \tau 1 1 } $ & $ 10^{-2} $ & $ \epsilon_{ \tau \tau 1 2 }  $ & $ 9.4 \times 10^{-6} $ & $ \epsilon_{ \tau \tau 1 3 } $ & $ 3.9 \times 10^{-3} $ \\
\hline
$ \epsilon_{ \tau \tau 2 2 } $ & $ 1.2 \times 10^{-1} $ & $ \epsilon_{ \tau \tau 2 3 } $ & $ 10^{-3} $ & $ \epsilon_{ \tau \tau 3 3 } $ & $ 8.6 \times 10^{-2} $ \\
\hline
\hline
$\epsilon_{ e \mu 1 1} $ & $ 8.5 \times 10^{-7} $ & $\epsilon_{ e \mu 1 2 } $ & $ 9.4 \times 10^{-6} $ & $\epsilon_{ e \mu 1 3 } $ & $ 3.9 \times 10^{-3} $ \\
\hline
$\epsilon_{ e \mu 2 1} $ & $ 9.4 \times 10^{-6} $ & $\epsilon_{ e \mu 2 2 } $ & $ 0.24 $ & $\epsilon_{ e \mu 2 3 } $ & $ 10^{-3} $ \\
\hline
$\epsilon_{ e \mu 3 1 } $ & $ 3.9 \times 10^{-3} $ & $\epsilon_{ e \mu 3 2 } $ & $ 10^{-3} $ & $\epsilon_{ e \mu 3 3 } $ & $ 6.6 \times 10^{-2} $ \\
\hline
\hline
$\epsilon_{ e \tau 1 1 } $ & $ 8.4 \times 10^{-4} $ & $\epsilon_{ e \tau 1 2 } $ & $ 9.4 \times 10^{-6} $ & $\epsilon_{ e \tau 1 3 } $ & $ 3.9 \times 10^{-3} $ \\
\hline
$\epsilon_{ e \tau 2 1 } $ & $ 9.4 \times 10^{-6} $ & $\epsilon_{ e \tau 2 2 } $ & $ 0.24 $ & $\epsilon_{ e \tau 2 3 } $ & $ 10^{-3} $ \\
\hline
$\epsilon_{ e \tau 3 1 } $ & $ 3.9 \times 10^{-3} $ & $\epsilon_{ e \tau 3 2 } $ & $ 10^{-3} $ & $\epsilon_{ e \tau 3 3 } $ & $ 0.2 $ \\
\hline
\hline
$\epsilon_{ \mu \tau 1 1 } $ & $ 9.4 \times 10^{-4} $ & $\epsilon_{ \mu \tau 1 2 } $ & $ 9.4 \times 10^{-6} $ & $\epsilon_{ \mu \tau 1 3 } $ & $ 3.9 \times 10^{-3} $ \\
\hline
$\epsilon_{ \mu \tau 2 1 } $ & $ 9.4 \times 10^{-6} $ & $\epsilon_{ \mu \tau 2 2 } $ & $ 0.24 $ & $\epsilon_{ \mu \tau 2 3 } $ & $ 10^{-3} $ \\
\hline
$\epsilon_{ \mu \tau 3 1 } $ & $ 3.9 \times 10^{-3} $ & $\epsilon_{ \mu \tau 3 2 } $ & $ 10^{-3} $ & $\epsilon_{ \mu \tau 3 3 } $ & $ 1 $ \\
\hline
\end{tabular}
\end{center}
\label{tab:2L2Q-YL}
\end{table*}
For the TLFV mediated by $S$, the last term in Eq.(\ref{eq:TLFV}), it is best  constrained by the neutral meson mixings.
Following the convention in \cite{UTfit},
the corresponding Wilson coefficients and the 4-fermi operators for $K$-$\bar{K}$, $B_d$-$\bar{B}_d$ and $B_s$-$\bar{B}_s$ mixing are
\begin{align}
&\tilde{C}^1_K=-\frac{1}{2m_S^2}(Y_s)_{11}(Y_s^\dagger)_{22},\quad \tilde{Q}^1_K=(\bar{s}\gamma^\mu\PR d)(\bar{s}\gamma_\mu\PR d)\,,\nonumber\\
&\tilde C^1_B=-\frac{1}{2m_S^2}(Y_s)_{11}(Y_s^\dagger)_{33},\quad \tilde{Q}^1_B=(\bar{b}\gamma^\mu\PR d)(\bar{b}\gamma_\mu\PR d)\,,\nonumber\\
&\tilde C^1_{B_s}=-\frac{1}{2m_S^2}(Y_s)_{22}(Y_s^\dagger)_{33},\quad \tilde{Q}^1_{B_s}=(\bar{b}\gamma^\mu\PR s)(\bar{b}\gamma_\mu\PR s)\,.
\end{align}
A global analysis with $95\%$ C.L. gave\cite{UTfit}
\begin{align}
& |\mathrm{Re}(\tilde C_K^1)|<9.6\times 10^{-13}, \quad
-4.4 \times 10^{-15}<\mathrm{Im}(\tilde C_K^1)<2.8 \times 10^{-15}\,,\nonumber\\
& | \tilde C_{B_d}^1 |<2.3 \times 10^{-11}, \quad
-\pi<\mathrm{Arg}(\tilde C_{B_d}^1)<\pi\,,\nonumber\\
& | \tilde C_{B_s}^1 |<1.1\times 10^{-9}, \quad
-\pi<\mathrm{Arg}(\tilde C_{B_s}^1)<\pi\,,
\end{align}
in the unit of $\mathrm{GeV}^{-2}$. Or equivalently,
\begin{align}\label{eq:constraint_on_yS}
&|(Y_s)_{11} (Y_s)^\dagger_{22}|< 1.92 \times 10^{-6} \times \left( \frac{m_S}{\text{TeV}} \right)^2 ,\nonumber\\
&|(Y_s)_{11} (Y_s)^\dagger_{33}|<  4.6 \times 10^{-5} \times \left( \frac{m_S}{\text{TeV}} \right)^2,\nonumber\\
&|(Y_s)_{22} (Y_s)^\dagger_{33}|<  2.2 \times 10^{-3} \times \left( \frac{m_S}{\text{TeV}} \right)^2.
\end{align}
For the democratic $Y_S$, the above constraints imply $|Y_S| \lesssim 9 \times 10^{-3} \times(m_S /7 \mbox{TeV})$.

Before ending this section, we recap the assumptions and discussion so far:
\begin{itemize}
\item  $Y_S$'s are assumed to be democratic and there is no outstanding hierarchy among these Yukawa couplings.
    This leads to one nearly massless active neutrino and $|Y_S| \lesssim 9 \times 10^{-3} \times(m_S /7 \mbox{TeV})$
    from the constrains of neutral meson mixings.
\item The Yukawa couplings $Y_R$ are turned off to minimize the TLFV.
\item For a given set of $\{\mu, m_S, m_\Delta, (Y_S)_{13,23,33}\}$ and any one of the $Y_L$'s, all the remaining 8 $Y_L$ can be iteratively determined from the absolute neutrino masses and the $U_{PMNS}$ matrix.
\end{itemize}

\section{Charged lepton flavor violating  process at one-loop}

\begin{figure*}[h]
\begin{center}
\includegraphics[width=16cm]{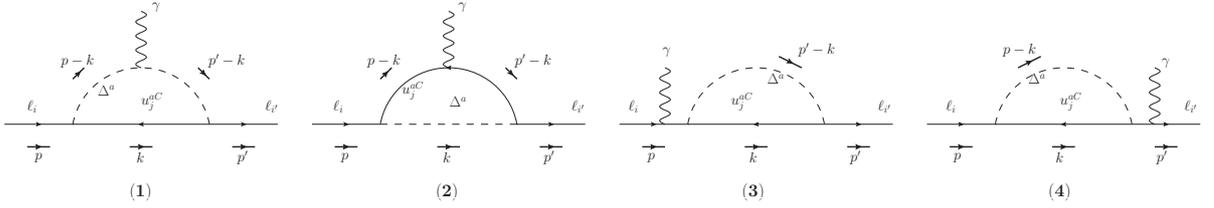}
\caption{The Feynman diagrams for 1-loop LFV $\mu\to e\gamma$.} \label{fig:meg}
\end{center}
\end{figure*}

In this section, we shall study the charged lepton flavor violating (cLFV) processes $\ell\to\ell' \gamma$, $Z\to \ell'\bar{\ell}$ and the like which are  induced at the 1-loop level with the leptoquark running in the loop, see Fig.\ref{fig:meg}.

\subsection{$\ell \to \ell'\gamma$}
The effective Lagrangian  responsible for the cLFV process $\ell\to \ell'\gamma$\cite{Kuno,WFC-LFV}  is parameterized as
\begin{equation}\label{meg-operator}
\mathscr{L} \supset \frac12 \bar{\ell}'\left( d_L^{ll'}\PL + d_R^{ll'} \PR\right)\sigma^{\mu\nu} \ell F_{\mu\nu} +h.c.
\end{equation}
For  $m_\ell' \ll m_\ell$, the partial decay width is given as
\beq
\Gamma(\ell \to \ell' \gamma) \simeq  { m_\ell^3 \over 16 \pi}  (|d^{ll'}_L|^2 + |d^{ll'}_R|^2)\,.
\eeq
A straightforward calculation yields
\beq
\label{eq:dR_MEG}
d_R^{ll'} =- {N_c e \over 16\pi^2 m_\Delta^2}\left[ \left( m_{l'}(Y^*_R)_{l'q}(Y^T_R)_{q l} +m_l (Y^*_L)_{l'q}(Y^T_L)_{q l}\right) {\cal F}_1(r_q) +m_q (Y^*_L)_{l'q}(Y^T_R)_{q l}{\cal F}_2(r_q)\right]\,,
\eeq
where the index $q$ sums over $q=u,c,t$ and $r_q\equiv m_q^2 /m_\Delta^2$.  $d_L^{ll'}$ can be obtained by simply switching $Y_L \leftrightarrow Y_R$ in the above expression for $d_R^{ll'}$.
The loop functions are
\beqa
{\cal F}_1(x) &=&{1+4x-5x^2+2x(2+x)\ln x \over 12(1-x)^4}\,,\nonr\\
{\cal F}_2(x)&=& {7-8x+ x^2 + 2(2+x)\ln x \over 6(1-x)^3}\,,
\eeqa
and they take the limits ${\cal F}_1\rightarrow 1/12$ and ${\cal F}_2\rightarrow 7/6+(2\ln x )/3$ when $x\rightarrow 0$.
Unlike at the tree-level, the contributions to the cLFV processes from $Y_L$ and $Y_R$ entangle with each other at the loop-level.
Since  $m_q {\cal F}_2(r_q) \gg m_l  {\cal F}_1(r_q)$ (for $q=c,t$),   generally speaking, the last term in Eq.(\ref{eq:dR_MEG}) which involves both $Y_L$ and $Y_R$ gives  the most important contribution to $d_R^{ll'}$\footnote{Barring the cases of fine-tuned cancelations and the hierarchical Yukawa couplings}. Therefore, it is expected that by setting $Y_R=0$  to minimize the TLFV
will also reduce the 1-loop cLFV processes in general. In the $Y_R=0$ case, $d_R^{ll'}$ dominates the cLFV processes because $m_\ell\gg m_{\ell'}$.
With $\tau_\mu^{-1}\approx \Gamma(\mu\to e\bar{\nu}_e\nu_\mu) = \frac{1}{192\pi^3}G_F^2 m_\mu^5$ and $\Gamma_\tau = 0.002265 \gev$,
the branching ratios for $\ell \to \ell' \gamma$ are
\beq
\mathcal{B}(\mu \to e\gamma)
\simeq  \frac{12\pi^2  |d_R^{\mu e}|^2 }{ G_F^2 m_\mu^2}\, ,\,\,
\mathcal{B}(\tau\to \ell' \gamma) \simeq  \frac{ m_\tau^3 |d_R^{\tau l'}|^2 }{16\pi \Gamma_\tau}\,.
\eeq
and
\beq
d_R^{ll'} = -\frac{e N_c m_\ell}{12 (4\pi)^2 m_\Delta^2}\sum_{q=u,c,t}  \left[  a_q^\gamma (Y_L)^*_{\ell' q} (Y_L)_{\ell q}   \right]\, ,
\eeq
where  $a^{\gamma}_q = 1 + 4 r_q (\ln r_q +1) +{\cal O}(r_q^2)$, $a^{\gamma}_u\sim a^{\gamma}_c\sim 1.0$ and $a^{\gamma}_t\sim 0.82$.
Numerically, we have
\begin{eqnarray}
\mathcal{B}(\mu \to e\gamma) &\simeq &   2.1 \times 10^{-7}\times \left|\sum_{q=u,c,t} a_q^{\gamma} (Y_L^*)_{eq} (Y_L)_{\mu q}  \right|^2\times\left( { 1\mbox{TeV} \over m_\Delta }\right)^4\, , \nonumber \\
\mathcal{B}(\tau \to \ell' \gamma) &\simeq & 3.8 \times 10^{-8}\times \left|\sum_{q=u,c,t} a_q^{\gamma} (Y_L^*)_{\ell' q} (Y_L)_{\tau q}  \right|^2 \times\left( { 1\mbox{TeV} \over m_\Delta }\right)^4\,.
\end{eqnarray}

\subsection{ Remark on other photon dipole induced processes}
\subsubsection{Anomalous magnetic dipole moment}
Similar calculation with little modification can be carried over  for the flavor diagonal cases. For the charged lepton,  the anomalous magnetic dipole moment is
\beq
 \triangle a_\ell = \frac{N_c }{6 (4\pi)^2} \frac{m_\ell^2}{m_\Delta^2} \left| \sum_{q=u,c,t} a_q^\gamma   (Y_L)_{\ell q} \right|^2\, .
\eeq
Assuming that $|(Y_L)_{l q}|^2\sim{\cal O}(1)$ and $m_\Delta=1$ TeV, one has $\triangle a_e\sim 8.0\times 10^{-16}$,  $\triangle a_\tau\sim 1.0\times 10^{-8}$,  and  $\triangle a_\mu\sim 3.0\times 10^{-11}$. Unless $m_\Delta\ll 1$TeV and all 3 $(Y_L)_{\mu q}$ are sizable and in phase, $\triangle a_\mu$ in this model is too small to accommodate the observed discrepancy
$ a_\mu^{exp}- a_\mu^{th} =(2.39\pm0.79)\times 10^{-9}$ at $1\sigma$ C.L.\cite{mu:g-2}.
Moreover, the model predicts a tiny positive $\triangle a_e$ which goes against the direction of the observed value that
$a_e^{exp}- a_e^{th} =-10.6(8.1)\times 10^{-13}$ at $1\sigma$ C.L.\cite{e:g-2}. Of course, a much larger $\triangle a_\mu$ is possible
to explain to observed discrepancy between the experimental measured value and the theoretical prediction if $Y_R\neq 0.$

\subsubsection{Electric dipole moments}
If $Y_R\neq 0$, the 1-loop  charged lepton  electric dipole moment(EDM),  $d_\ell \sim \frac{N_c}{16\pi^2} \frac{m_t}{m_\Delta^2} \mathbf{Im}[Y_L Y_R^*]$, could be large. For $m_\Delta=1 \tev$, $|Y_L|\sim |Y_R|\sim 0.01$, and the CP phase is of order one, the typical electron EDM is around $10^{-24}\, e$-cm which is already 4 orders of magnitude larger than the current limit $|d_e|<8.7\times 10^{-29} e$-cm \cite{e:edm}.
Then, how to suppress the EDM's in this model will be a pressing theoretical issue. A plain solution is setting $m_\Delta \gtrsim 100 \tev$ to avoid the too large EDMs but the phenomenology at the low energies are strongly suppressed as well.

On the other hand, if $Y_R=0$ there is no EDM at the 1-loop level.
In fact, the first non-zero EDM contribution we can construct begins at the 3-loop level and it  involves both $V^{CKM}$ and $U^{PMNS}$. An order of magnitude estimate  gives:
\beq
d_\ell \sim \frac{\alpha N_c}{(16\pi)^3} \frac{m_\ell}{m_\Delta^2} \mathbf{Im}\left[(Y_L)_{\ell k} V^{CKM}_{kj}(Y_L^\dag)_{ji}U^{PMNS}_{i\ell}\right]\,.
\eeq
If $Y_L$ takes a typical value $\sim 0.01$, $m_\Delta=1 \tev$, and the combined  CP phase is $\sim{\cal O}(1)$, this 3-loop electron EDM is expected to be $|d_e|\lesssim  10^{-37}\,e$-cm.
This upper bound is slightly larger than the estimated SM upper bound for $d_e$ but way below the sensitivity of any  EDM measurement in the foreseeable future.
Consequently, $d_e$ is a useful handle to test the $Y_R=0$ assumption in the cZBM:
once the electron EDM was measured to be greater than $10^{-37}\, e$-cm,  either  the $Y_R=0$ assumption with $m_\Delta \sim {\cal O}(\tev)$ must be  abandoned or more new physics is needed to go beyond the cZBM.

\subsubsection{$\mu-e$ conversion}
The $\mu-e$ conversion(MEC) will be mediated by the leptoquark at the tree-level as shown in Fig.\ref{fig:TL_FV}. For $Y_R=0$, the relevant 4-fermi operator is
\beq
 \frac{(Y_L)^*_{e u} (Y_L)_{\mu u}  }{2 m_{\Delta}^2} \overline{e} \gamma^{\mu} \PP_{L} \mu \cdot \overline{u^{a}} \gamma_{\mu} \PP_{L} u^{a}+h.c.
\eeq
The cLFV photon dipole operator  discussed in the previous section will also contribute to MEC with an expected  relative magnitude $\sim (\alpha/16\pi^2)^2$ comparing to the tree-level one.
Following the analysis of \cite{WFC-LFV}, a more quantitative estimate for the MEC rate is:
\begin{eqnarray}
\mathcal{B}_{conv}  &\simeq & \mathcal{C}_{conv} \left\{ 	  \left( \frac{\alpha  Z}{16\sqrt{2}\pi G_F m_\Delta^2} \right)^2 |\sum_q a^\gamma_{q}(Y_L)^*_{eq} (Y_L)_{\mu q}|^2	 \right.  \nonumber \\
				   && \left. +  \left( \frac{(2Z+N)}{4 \sqrt{2} G_F m_\Delta^2} \right)^2  |(Y_L)^*_{eu} (Y_L)_{\mu u}|^2   \right\} \nonumber \\
				   &\simeq & \mathcal{C}_{conv} \left\{ 8.9 \times 10^{-11} Z^2	|\sum_q a^\gamma_{q}(Y_L)^*_{eq} (Y_L)_{\mu q}|^2 \right. \nonumber \\
				    && \left. +  2.3 \times 10^{-4} (2Z+N)^2 |(Y_L)^*_{eu} (Y_L)_{\mu u}|^2  \right\} \times \left[ \frac{\tev}{m_\Delta} \right]^4 \ ,
\end{eqnarray}
where Z is the atomic number and N is the neutron number for a certain nucleus.
The overall factor $\mathcal{C}_{conv}$  depends on the form factors of the nuclei and the momentum of the muon.
For instance,  $\mathcal{C}_{conv} ({}^{48}_{22}Ti) = 1.2\times 10^{-3}$\cite{2L2Q}.
As can be seen, the LFV photon dipole indeed has much smaller contribution to the MEC than the tree-level one.

\subsubsection{$\mu\rightarrow 3e$  }
In this model, there are no tree-level contributions to the cLFV  $\mu\rightarrow 3e$ decay.
The $\mu\rightarrow 3e$ process is dominated by the cLFV photon dipole transition and its rate is much smaller than ${\cal B}(\mu\rightarrow e \gamma)$.
As pointed out in \cite{Kuno,WFC-LFV}, the ratio of  ${\cal B}(\mu\rightarrow 3e)$ to ${\cal B}(\mu\rightarrow e \gamma)$  is basically  model-independent:
\beq
{{\cal B}(\mu\rightarrow 3e) \over {\cal B}(\mu\rightarrow e \gamma)}\sim \frac{2\alpha}{3\pi} \left[\ln \frac{m_\mu}{m_e}-\frac{11}{8}\right]\simeq 0.019\,.
\eeq
Similarly, with replacing the charged lepton masses, the ratios in the rare tau decays are
\beqa
{{\cal B}(\tau\rightarrow 3e) \over {\cal B}(\tau\rightarrow e \gamma)}\sim \frac{2\alpha}{3\pi} \left[\ln \frac{m_\tau}{m_e}-\frac{11}{8}\right]\simeq 0.011\,,\nonr\\
{{\cal B}(\tau\rightarrow 3\mu) \over {\cal B}(\tau\rightarrow \mu \gamma)}\sim \frac{2\alpha}{3\pi} \left[\ln \frac{m_\tau}{m_\mu}-\frac{11}{8}\right]\simeq 0.002\,.
\eeqa
For the decay channels with different flavor final sates, one has\cite{WFC-LFV}
\beqa
{{\cal B}(\tau\rightarrow \mu e e^+ ) \over {\cal B}(\tau \rightarrow e \gamma)}\sim \frac{2\alpha}{3\pi} \left[\ln \frac{m_\tau}{m_e}-\frac{3}{2}\right]\simeq 0.032\,,\\
{{\cal B}(\tau\rightarrow e \mu \mu^+) \over {\cal B}(\tau\rightarrow e \gamma)}\sim \frac{2\alpha}{3\pi} \left[\ln \frac{m_\tau}{m_\mu}-\frac{3}{2}\right]\simeq 0.0064\,.
\eeqa
The decay branching ratios $\tau\rightarrow \mu^+ e e $ and $\tau\rightarrow \mu \mu e^+ $ are negligible because they are doubly suppressed by two cLFV transition vertices.

%%%%%%%%%%%%%%%%%%%%%%%%%%%%%%%%%%%%%%%%%%%%%%%%%%%%%%%%%%%%%%%%%%%%%%%%%%%%%%
%				Z to l-bar l'
%%%%%%%%%%%%%%%%%%%%%%%%%%%%%%%%%%%%%%%%%%%%%%%%%%%%%%%%%%%%%%%%%%%%%%%%%%%%%%
\subsection{$Z \to \bar{\ell} \ell'$}
The same Feynman diagrams in Fig.\ref{fig:meg} with photon replaced by $Z$ boson lead to cLFV $Z\rightarrow \bar{l}l'$ decays.
Since $Z$ is massive, it can also admit the vector or axial-vector couplings other than the dipole transition couplings as in the $l\rightarrow l'\gamma$ cases.
The most general gauge invariant $Z\rightarrow \bar{l} l'$ amplitude is parameterized as:
\begin{equation}
i\mathscr{M}=ie\overline{u}(p') \left[ (c_R^Z\PR+c_L^Z\PL)\left(-g_{\mu\nu}+\frac{q_\mu q_\nu}{m_Z^2}\right)\gamma^\nu
+\frac{1}{m_Z}\left(d_L^Z\PL+d_R^Z\PR\right)(i\sigma_{\mu\nu}q^\nu)\right]v(-p)\epsilon^\mu(q)\,,
\end{equation}
where the 4-momentums are labeled as in Fig.\ref{fig:meg}.
From the above parametrization, the branching ratio can be easily calculated to be
\begin{eqnarray}
\mathcal{B}(Z\to\overline{\ell}{\ell}')=\frac{\alpha}{6}\frac{m_Z}{\Gamma_Z} \left[(|c_L^Z|^2+|c_R^Z|^2)+\frac{1}{2}\left(|d_L^Z|^2
+|d_R^Z|^2\right)\right] \ ,
\end{eqnarray}
and the experimentally measured value $\Gamma_Z=2.4952\pm0.0023\mathrm{GeV}$\cite{PDG} is used in our study.

The 4 dimensionless coefficients $c^Z_{R,L}, d^Z_{R,L}$ can be obtained through a lengthy but straightforward calculation. The physics is rather simple and can be understood qualitatively. However, the full analytic results  are not very illustrating and will not be presented here\footnote{The details will be given in other place.}.
Let's focus on the $Y_R=0$ case to simplify the physics discussion.
 First of all, the masses of external charged leptons are much smaller than $m_Z$ and they can be treated massless\footnote{Since  $m_u, m_c \ll m_Z$, they can also be treated as massless particles in this process.}.
 For $c_R^Z$, the coupling connects both left-handed fermions and there is no need to flip their chiralities. In the loop calculation, $m_Z$ and $m_t$ are the only two dimensionful quantities  other than  $m_\Delta$. So, by dimensional analysis we know that $c_R^Z \sim {\cal O}(\frac{N_c}{16\pi^2}\frac{m_t^2}{m_\Delta^2})$ (for top quark running in the loop) or $c_R^Z \sim {\cal O}(\frac{N_c}{16\pi^2}\frac{m_Z^2}{m_\Delta^2})$( for light quarks running in the loop).
For the dipole couplings which connect fermions with different handiness,  one external charged lepton mass insertion is needed to flip its chirality.
Also, $m_Z$ sets the nature scale of the
momentum transfer in this process.  Therefore it is expected that
in general $d^Z/c^Z_R \sim {\cal O}(\frac{m_l}{m_Z})$ or ${\cal O}(\frac{m_l'}{m_Z})$. Thus, the contributions from $d_{L,R}^Z$  can be safely ignored.
 On the other hand,  both of the two external charged leptons need to flip their chiralities for having a nonzero $c_L^Z$ if $Y_R=0$. Therefore, $c_L^Z \sim {\cal O}(\frac{m_l m_l'}{m_Z^2}) c_R^Z$ and its contribution is totally negligible in this process. The above qualitative understandings agree very well with our full calculation.
Hence, only the leading contribution from $c^Z_R$ is kept in the study.
It is more useful to express the final result in the numerical form:
\beq
\label{eq:BZllp}
\mathcal{B}(Z\to \bar{\ell} \ell')  \simeq  1.46 \times 10^{-7} \left|\sum_{q=u,c,t}a^Z_q (Y_L)^*_{\ell' q} (Y_L)_{\ell q}\right|^2
\times \left( \frac{\tev}{m_\Delta} \right)^4\, ,
\eeq
where $a_u^{Z} = a_c^{Z} \simeq -0.125-0.077 \mathbf{i} =-0.1468 e^{ i 31.63^\circ}$ and $a_t^{Z} = 1$.
The imaginary part of $a^Z_{u,c}$ comes from the pole of light-quark propagators in the loop when the light quarks  are going on-shell in the $Z$ decay.
Also note that this cLFV decay branching ratio is around $10^{-7}$ if the absolute square in Eq.(\ref{eq:BZllp}) is of order unit.
The ballpark estimate is  below  but close to the current experimental limits\cite{PDG,ATLAS-Zdecay-2014}.

The interference between the sub-diagrams with $u(c)$ and $t$ running in the loop makes the relative phases between $a_{u,c}^Z$ and $a_t^Z$
observable. This physical phase leads to CP violation and in general $\mathcal{B}(Z \to \bar{\ell} \ell') \neq \mathcal{B}( Z \to \bar{\ell'} \ell)$. Following\cite{ZCP_Bernabeu,ZCP_Rius}, the CP asymmetries are quantified as:
\beq
\eta_{\ell \ell'} \equiv  \mathcal{B}(Z \to \bar{\ell} \ell') - \mathcal{B}(Z \to \bar{\ell'} \ell)\,.
\eeq
In this model, we have  numerically
\beq
\eta_{\ell \ell'}\simeq  (4.53 \times 10^{-8} ) \times {\bf Im}\left[ \left(  Y^{\ell'\ell}_u + Y^{\ell' \ell}_c  \right) (Y^{\ell' \ell}_t)^* \right] \times \left( \frac{\tev}{m_\Delta} \right)^4\,,
\eeq
where the shorthand notation $Y^{\ell' \ell}_q \equiv (Y_L)^*_{\ell' q} (Y_L)_{\ell q}$. Interestingly, due to the sizable CP phase, the CP asymmetries and the cLFV decay branching ratios are of the same order. Also, for the later convenience, we define
$ \mathcal{B}^Z_{ \ell \ell'} \equiv \mathcal{B}(Z\to \bar{\ell} \ell') + \mathcal{B}(Z\to \ell \bar{\ell'})$.

Before closing this section, we should point out a simple but useful scaling relationship between $Y_S$ and $Y_L$ in this model. Recall that the neutrino mass is proportional to $Y_S Y_L^2$. Therefore, if  $Y_S$ is re-scaled by $Y_S \rightarrow \lambda^{-2} Y_S$, then $Y_L$ must goes like $Y_L \rightarrow \lambda Y_L$ to keep the neutrino mass unchanged. After such rescaling, ${\cal B}(\ell\rightarrow \ell'\gamma)$, MEC,  $\mathcal{B}^Z_{ \ell \ell'}$ and $\eta_{\ell\ell'}$ go like $\lambda^{4}$ while $\epsilon_{ll'qq'}$, $\Delta a_l$, and EDM go like $\lambda^{2}$ due to their amplitude nature. This scaling relationship largely helps   reduce the computer time in finding the realistic configurations.

Now we have everything needed for the numerical and phenomenological study.

%%%%%%%%%%%%%%%%%%%%%%%%%%%%%%%%%%%%%%%%%%%%%%%%%%%%%%%%%%%%%%%%%%%%%%%%%%%%%%
%				numerical setup and study
%%%%%%%%%%%%%%%%%%%%%%%%%%%%%%%%%%%%%%%%%%%%%%%%%%%%%%%%%%%%%%%%%%%%%%%%%%%%%%
\section{Numerical Study}\label{sec: numerical study}

\subsection{Scanning strategy}
As discussed in Sec.3, once the set $\{\mu, m_S, m_\Delta, (Y_S)_{13,23,33}\}$ plus any one out of the 9 $Y_L$'s are fixed, all the remaining 8 $Y_L$'s can be iteratively determined from the absolute neutrino masses and the $U_{PMNS}$ matrix.
In our numerical search, we take  $m_\Delta = 1\tev$ and $m_S = 7 \tev$ as the benchmark.
Moreover,  for each configuration,  $\mu$ is randomly produced within $[0.1 ,1]\tev$.
For each search, the neutrino mixings $sin^2\theta_{12,13,23}$,  and the Dirac phase  $\delta_{cp}$ are randomly generated within the 1 sigma allowed range from the global fit, Tab.\ref{tab:nu_exp}. For simplicity the two Majorana phases are set to be zero. Then the $U_{PMNS}$ matrix can be determined via Eq.(\ref{eq:definition_of_pmns}).
For a given $U_{PMNS}$, we still need to know the absolute neutrino eigen-masses in order to obtain  the neutrino mass matrix,  see Eq.(\ref{eq:mnu_diag}).
As has been discussed, we assume the lightest neutrino mass is zero. Depending on the neutrino mass hierarchy, the other 2 absolute neutrino masses can be determined  from the given $\Delta m_{21}^2$ and $\Delta m^2$.
These 2 mass squared differences are also randomly generated within the 1 sigma allowed range from the global fit. Then the absolute neutrino mass matrix $M_\nu^{IH}(M_\nu^{NH})$ for the inverted(normal) hierarchy is ready for use.

Next, $|(Y_L)_{13}|$ is randomly generated as a real number between $10^{-10}$ and $1.0$.
Because of the scaling relationship discussed in the previous section,  we fix $|(Y_S)_{33}|=0.0097$\footnote{We have $|(Y_S)_{11}|\sim|(Y_S)_{22}|\sim|(Y_S)_{33}|$ and the most stringent bound is $|(Y_S)_{11} (Y_S)_{22}| < 9.408 \times 10^{-5}$, hence $|(Y_S)_{33}| \lesssim  0.0097 $, for $m_S = 7 \tev$.} without losing any generality.
Then, $|(Y_S)_{13,23}|$ are generated within $[0.1,10]\times 0.0097$ and they  must obey $ 0.1<|(Y_S)_{13}/(Y_S)_{23}| <10.0$ to be consistent with our working assumption.
The signs of $(Y_L)_{13}$ and $(Y_S)_{13,23,33}$ are also randomly assigned with equal probabilities being positive or negative.
With the above mentioned values, $(Y_L)^{(0)}_{23,33,12,22,32}$ can be fixed via Eq.(\ref{eq:YL_leading}).
Finally, $(Y_L)^{(1)}_{11,21,31}$ can be derived from the next order perturbation, Eq.(\ref{eq:YL_next}).
For that, we put in a small random perturbation within the range that $ 10^{-7}<|\delta_{23}/(Y_S)_{23}|, |\delta_{33}/(Y_S)_{33}| <10^{-2}$.

With all $Y_L$'s ready, the randomly generated  configuration is further checked to see whether it is viable.
A configuration will be accepted if it pass all the following criteria:
\begin{itemize}
  \item All  $|Y_L|$'s are less than one  so that the model can be calculated perturbatively.
  \item All the TLFV satisfy the current experimental limits listed in Tab.\ref{tab:2L2Q-YL}.
  \item All the loop-level cLFV processes must comply with the latest experimental limits\footnote{While we are wrapping up this article, the MEG Collaboration has updated the $\mathcal{B}(\mu\to e\gamma)$  limit with a slightly better value $<4.2\times 10^{-13},\; 90\%$ C.L.\cite{MEG2016}. } summarized in Tab.\ref{tab:loopLFV_exp}.
\begin{table*}[h]
\caption{
Summary of the latest experimental limits we used in the numerical scan.}
\begin{center}
\begin{tabular}{c|c}
\hline\hline
$\mathcal{B}(\mu^+\to e^+\gamma)$  & $<5.7 \times 10^{-13},\; 90\%$ C.L. \cite{MEG}\\
$\mathcal{B}(\tau\to \mu\gamma)$  & $<4.4 \times 10^{-8},\; 90\%$ C.L. \cite{BaBar:tau-LFV} \\
$\mathcal{B}(\tau\to e\gamma)$  & $<3.3 \times 10^{-8},\; 90\%$ C.L. \cite{BaBar:tau-LFV} \\
\hline\hline
$\mathcal{B}^Z_{\tau\mu}$  & $<1.2\times 10^{-5}$,\; $95\%$ C.L. \cite{PDG} \\
$\mathcal{B}^Z_{\tau e}$  &  $<9.8\times 10^{-6}$,\; $95\%$ C.L. \cite{PDG} \\
$\mathcal{B}^Z_{\mu e}$  &  $<7.5\times 10^{-7}$,\; $95\%$ C.L. \cite{ATLAS-Zdecay-2014} \\
\hline\hline
%$\Delta a_\mu$  & $(2.39\pm0.79)\times 10^{-9}\, (1\sigma)$ \cite{mu:g-2} \\
%$\Delta a_e$  &  $-10.6(8.1)\times 10^{-13}\, (1\sigma)$ \cite{e:g-2} \\
%\hline\hline
\end{tabular}
\end{center}
\label{tab:loopLFV_exp}
\end{table*}

\end{itemize}
The phenomenologically viable configurations are collected and then used to calculate the resulting cLFV.

\subsection{Numerical Result}

\subsubsection{$\ell \to \ell' \gamma$ and $Z\to \overline{\ell} \ell'$}
\begin{center}
\begin{figure}[h]
  \subfigure[]{\includegraphics[width=0.3\textwidth]{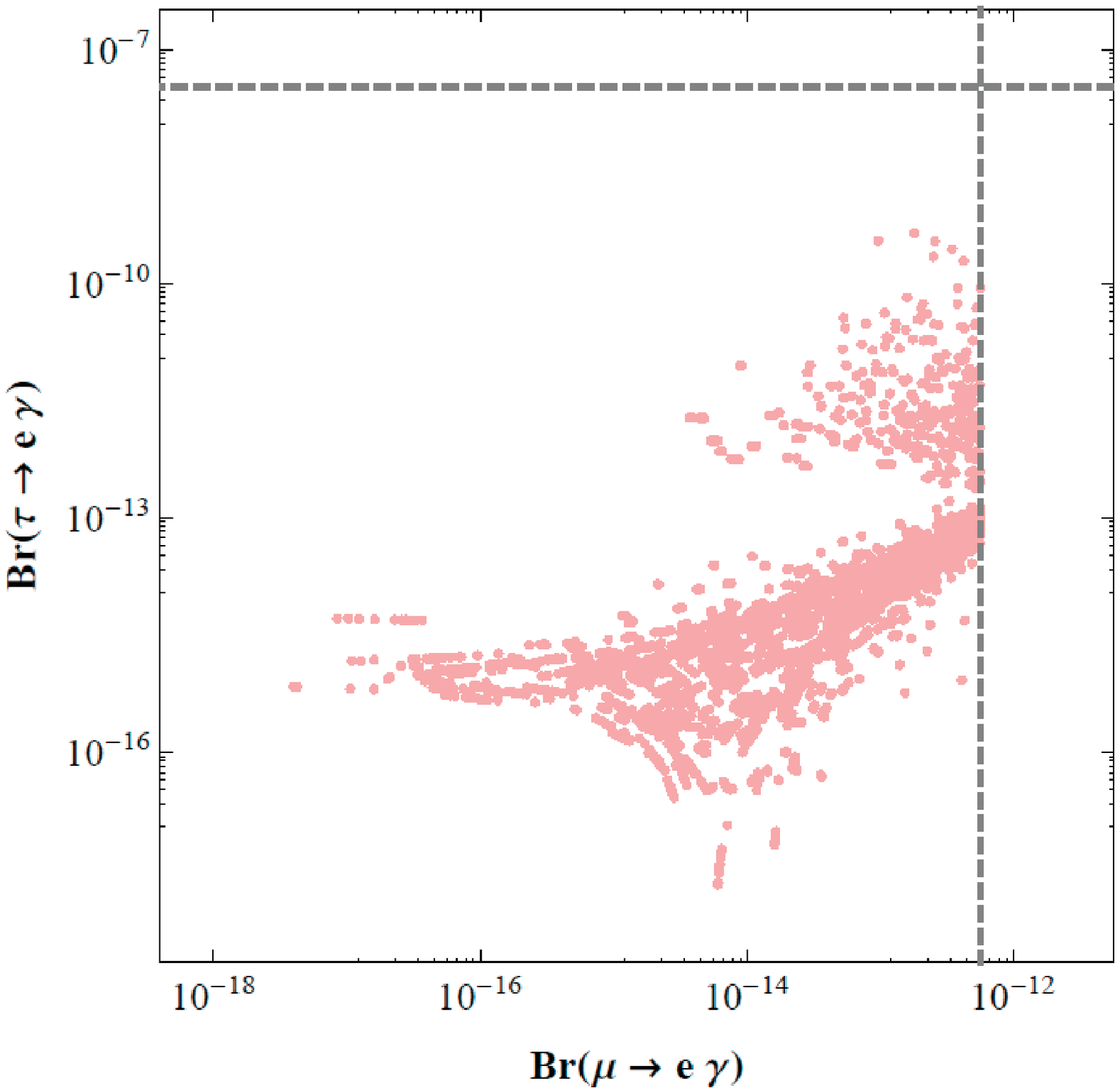}}
  \subfigure[]{\includegraphics[width=0.3\textwidth]{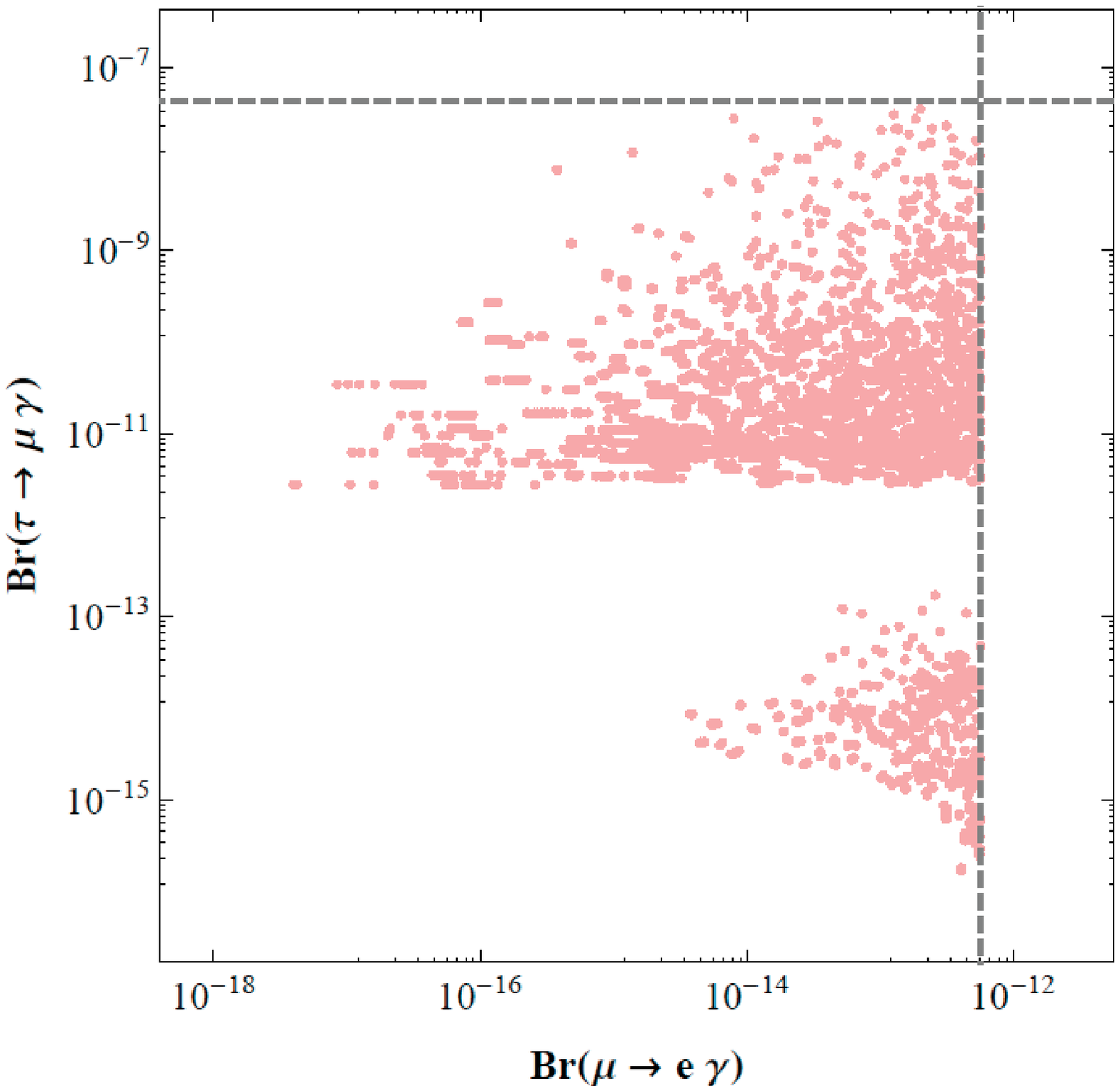}}
  \subfigure[]{\includegraphics[width=0.3\textwidth]{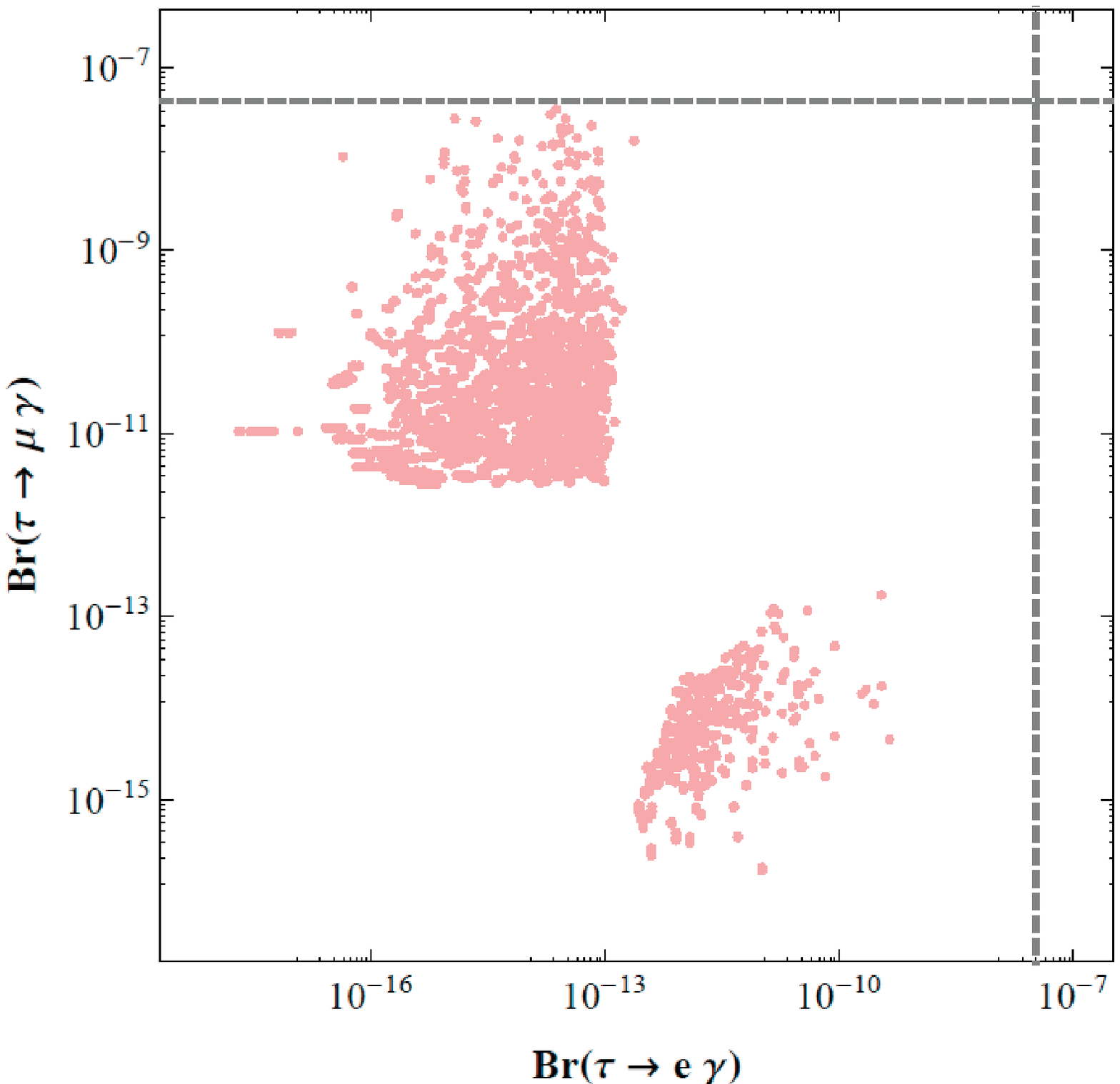} }\\
  \subfigure[]{\includegraphics[width=0.3\textwidth]{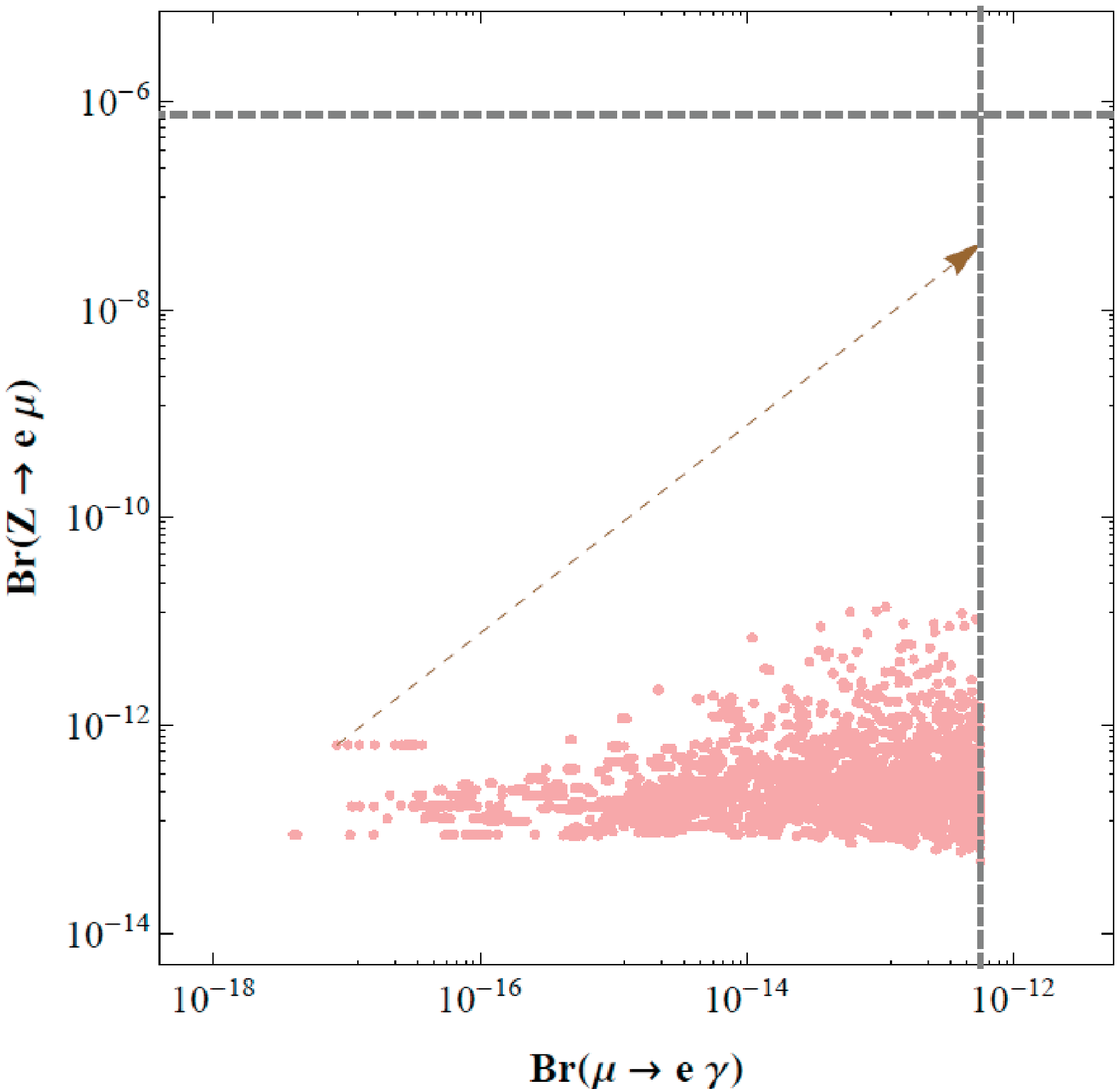} }   \subfigure[]{\includegraphics[width=0.3\textwidth]{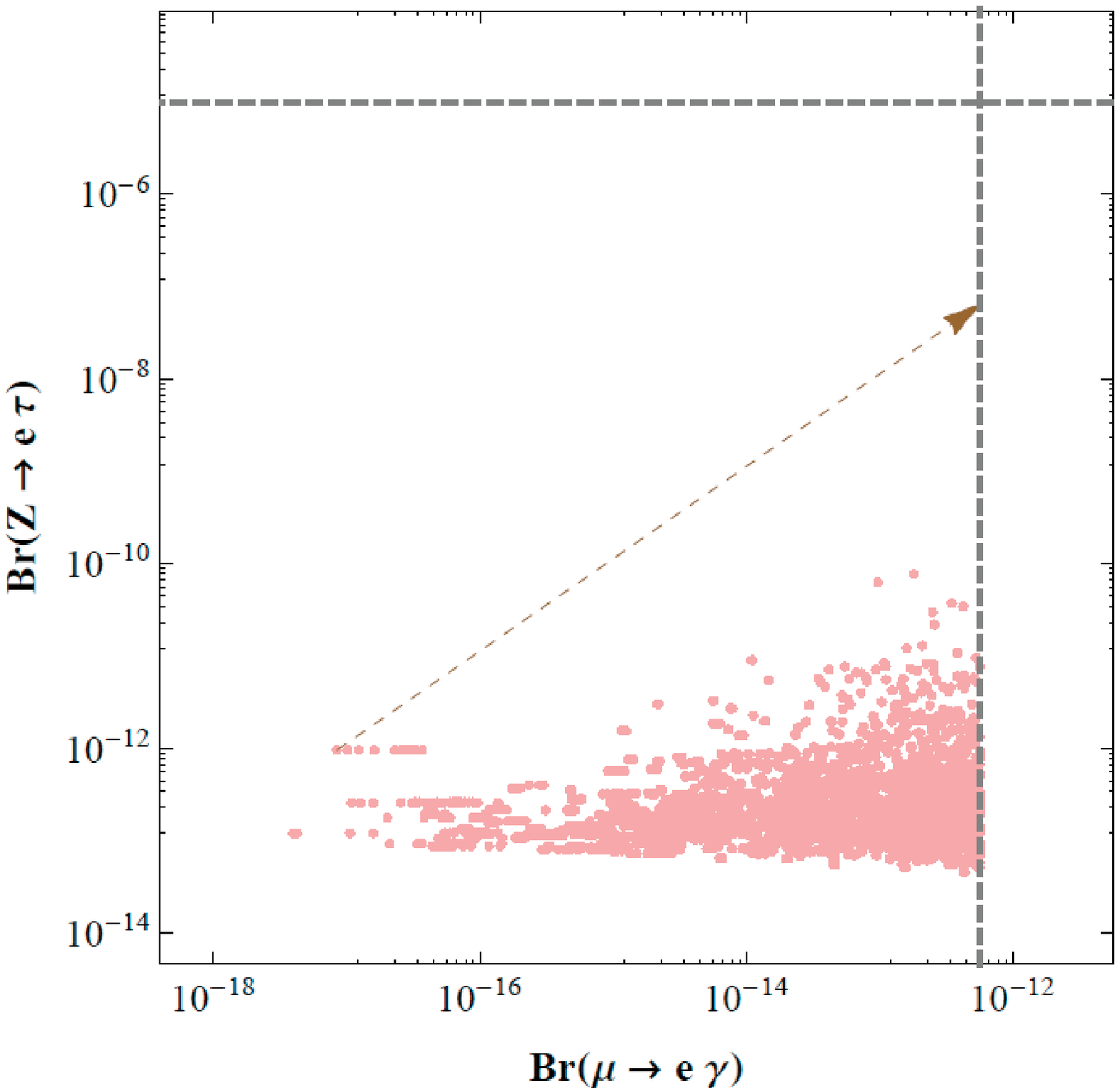} }
  \subfigure[]{\includegraphics[width=0.3\textwidth]{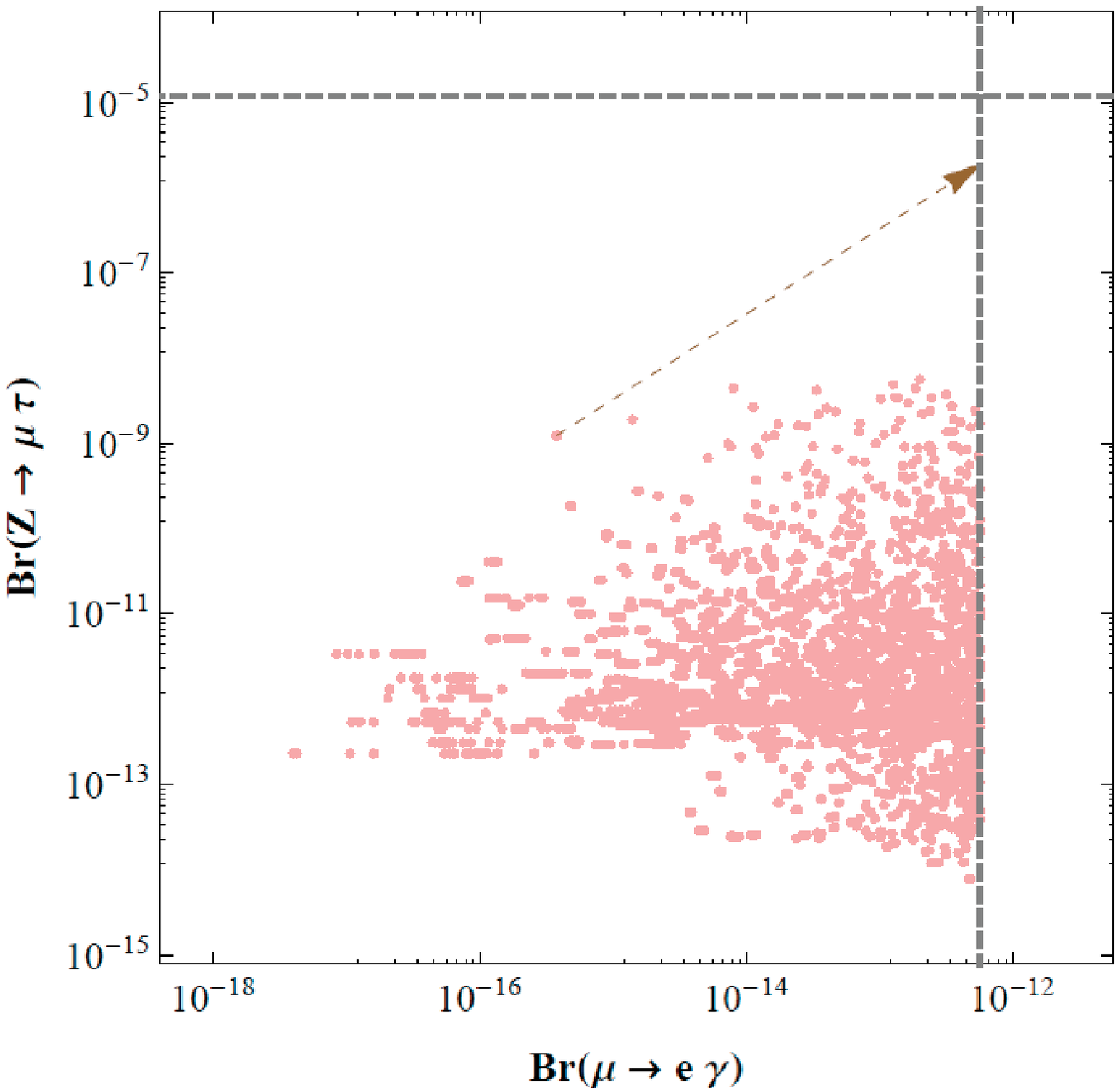}}
\caption{Correlations among the charged lepton flavor violating branching ratios  for neutrino masses are of the inverted hierarchy.
In these plots, $m_\Delta=1 \tev$, $m_S=7\tev$ and $|(Y_S)_{33}| =0.0097$.  The dashed lines represent the current experimental limits at 90\%C.L. }
\label{fig:IH_LFV}
\end{figure}
\end{center}

\begin{center}
\begin{figure}[h]
  \subfigure[]{\includegraphics[width=0.3\textwidth]{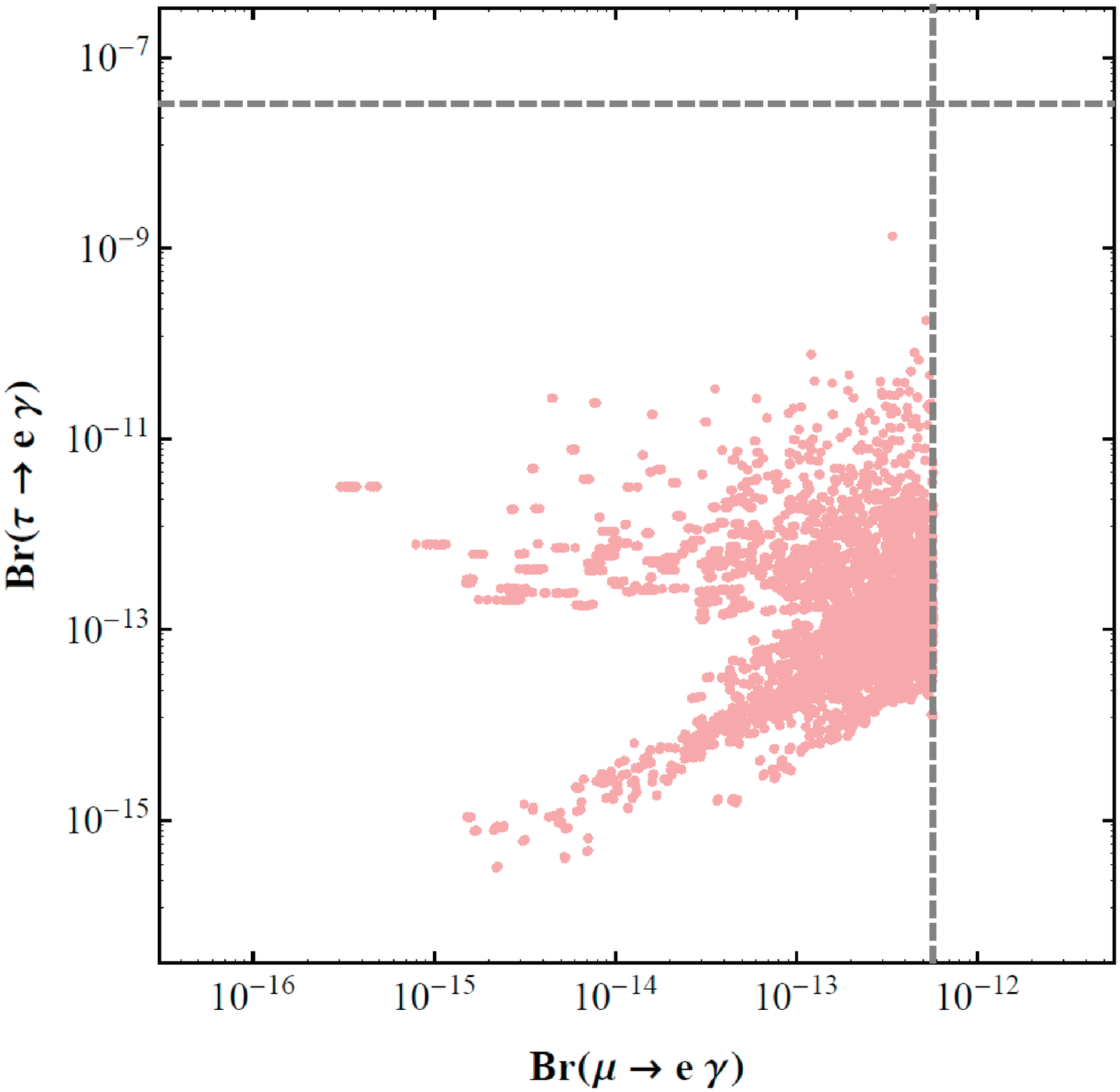}}
  \subfigure[]{\includegraphics[width=0.3\textwidth]{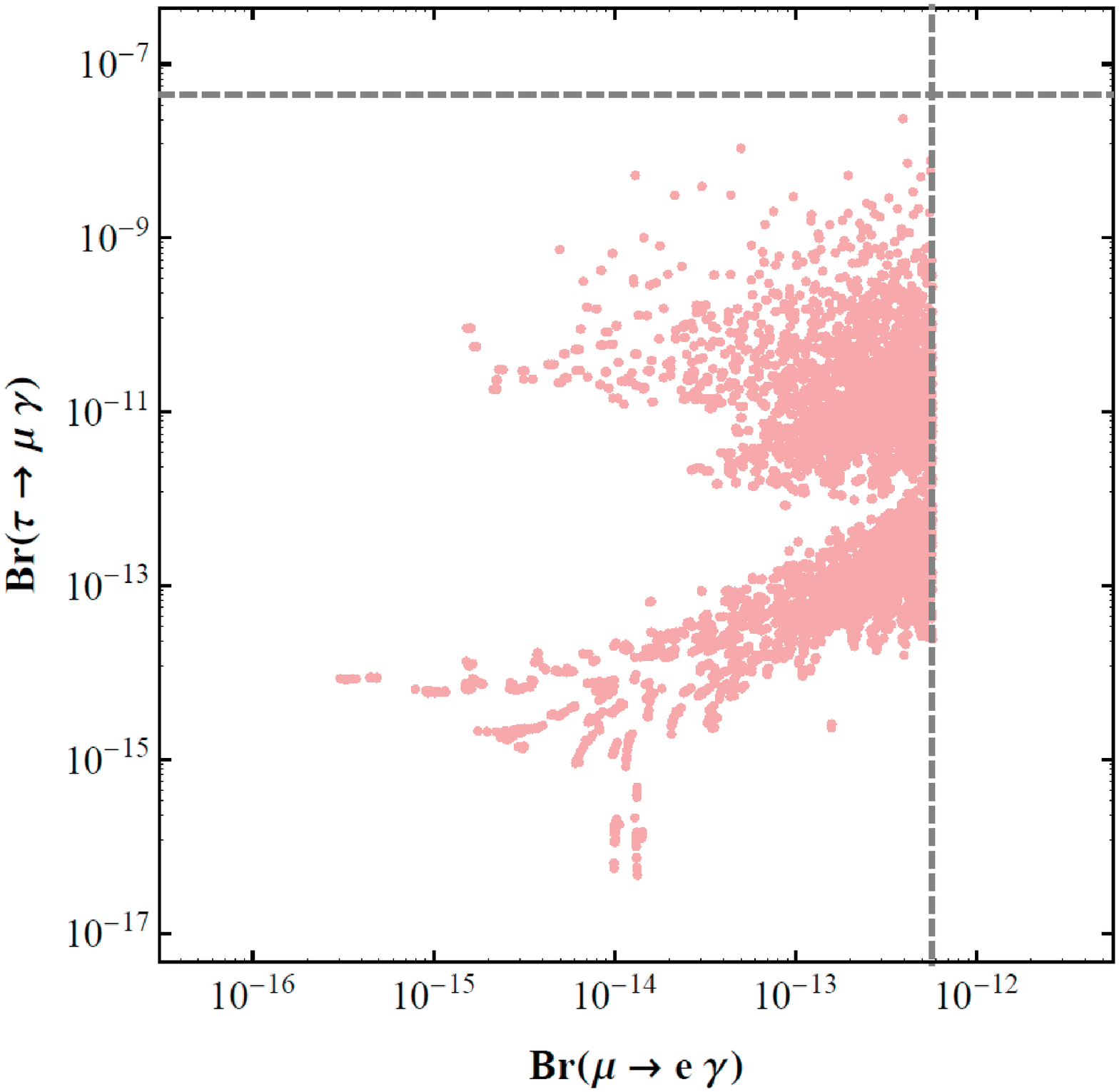}}
  \subfigure[]{\includegraphics[width=0.3\textwidth]{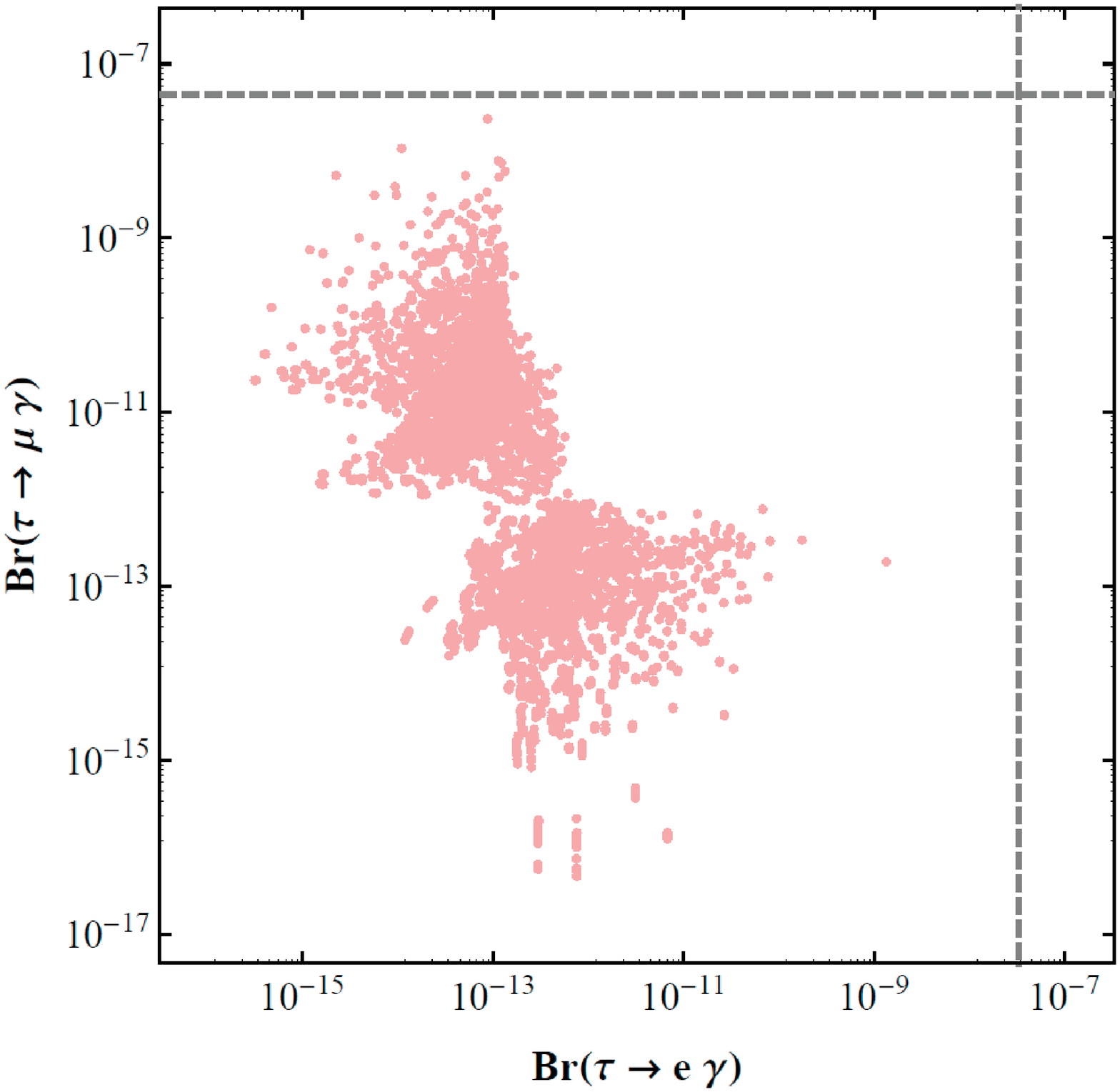}}\\
  \subfigure[]{\includegraphics[width=0.3\textwidth]{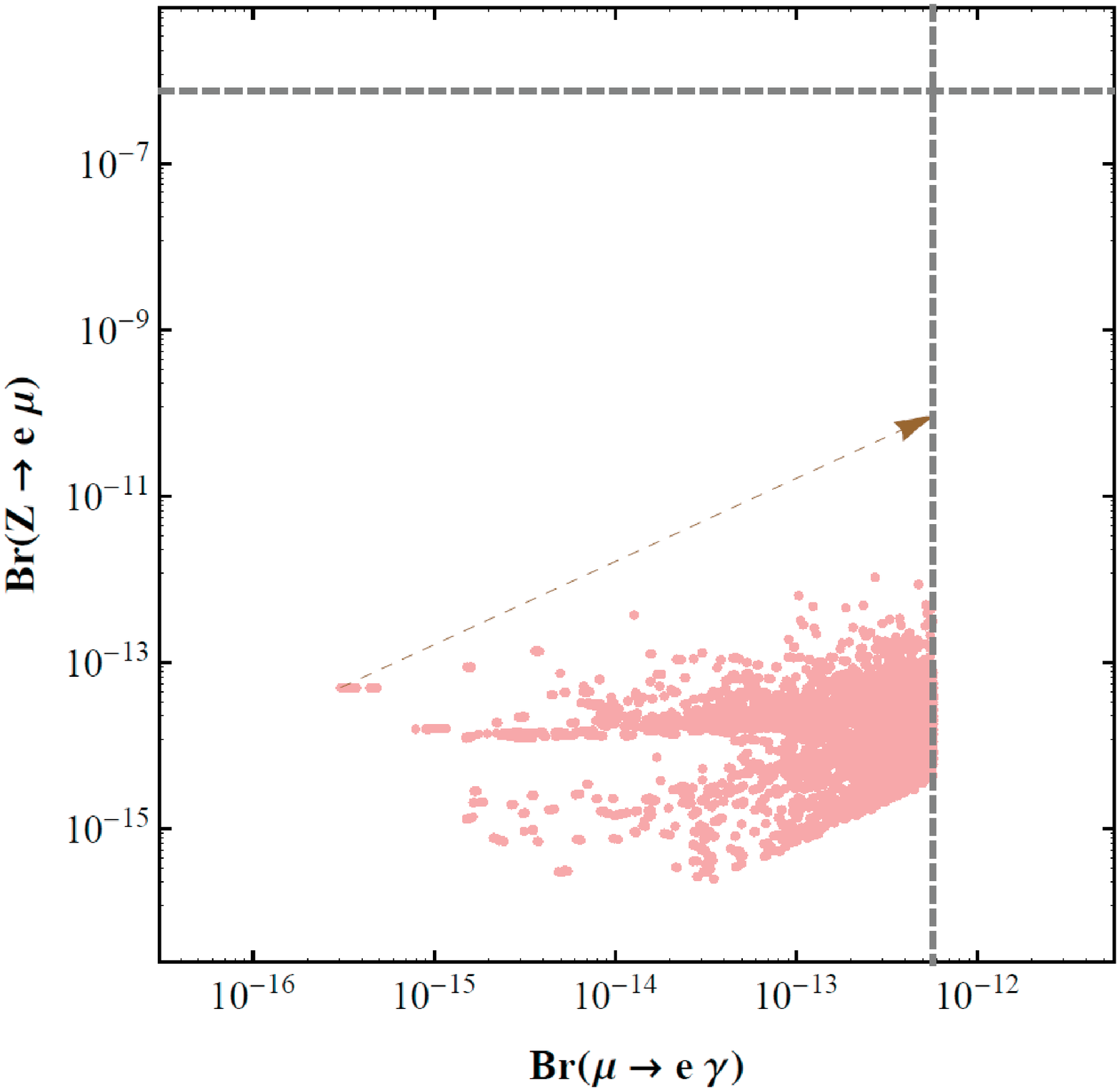}}
  \subfigure[]{\includegraphics[width=0.3\textwidth]{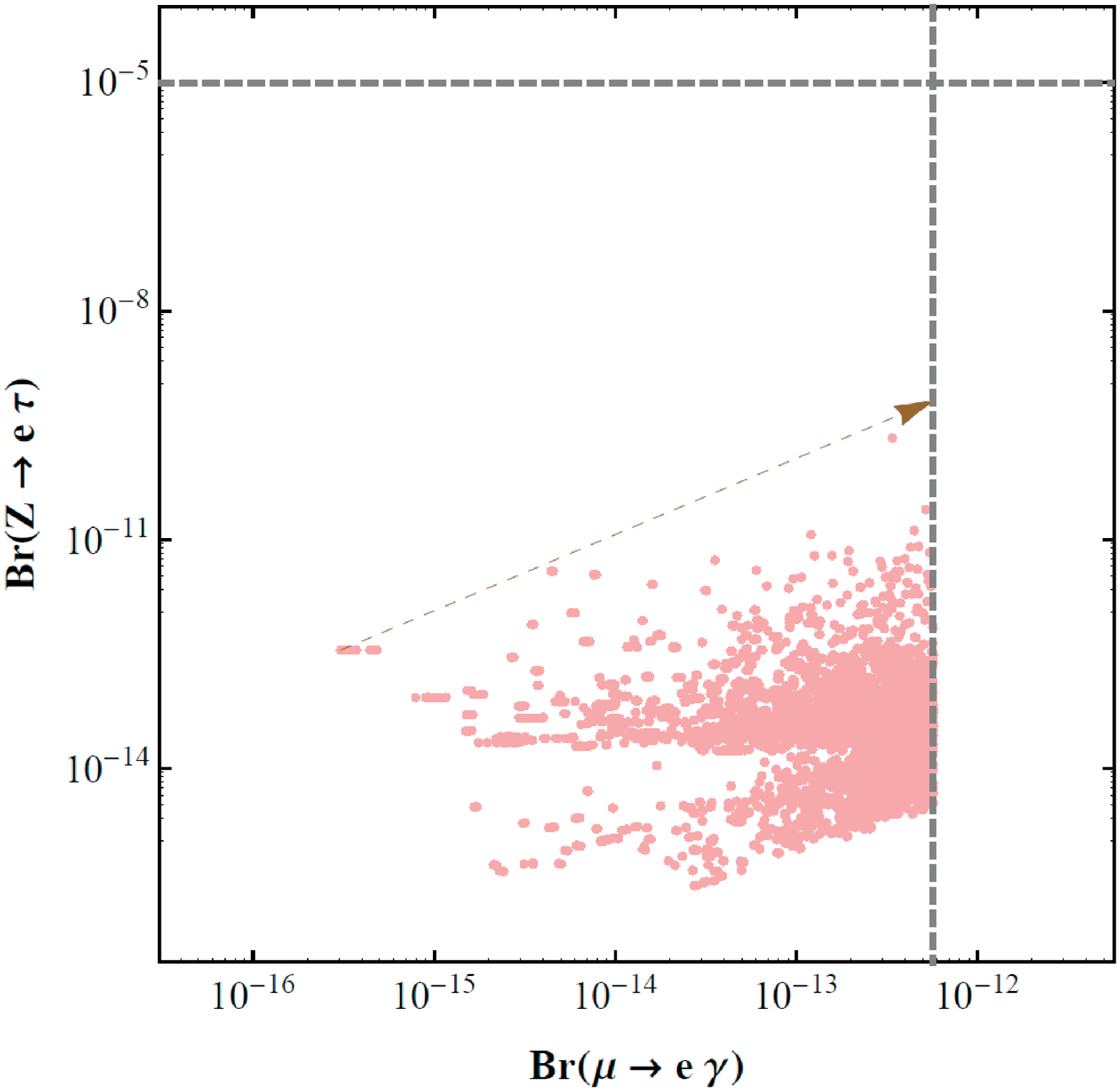}}
  \subfigure[]{\includegraphics[width=0.3\textwidth]{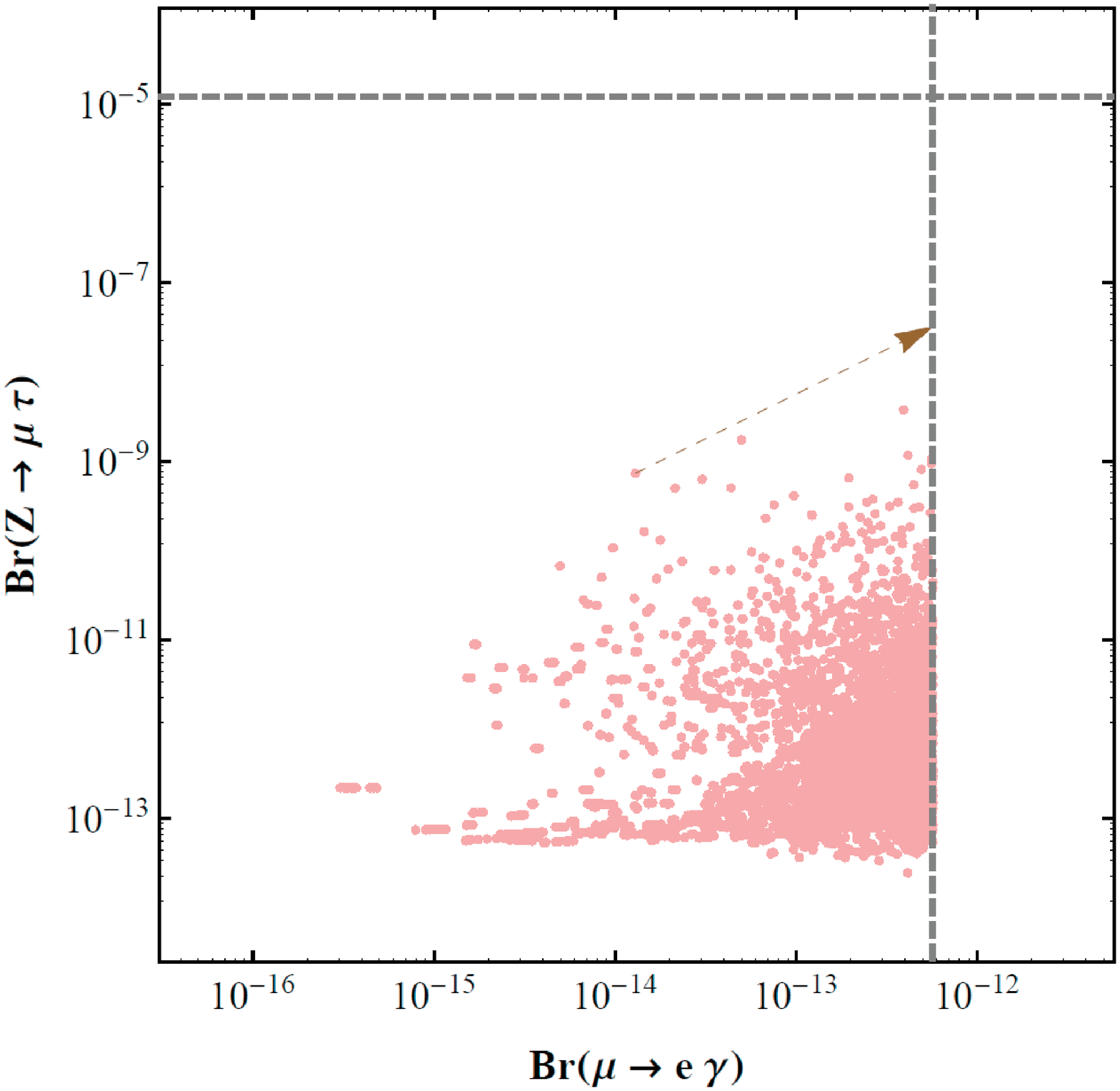}}
\caption{Correlations among the charged lepton flavor violating branching ratios for neutrino masses are of the normal hierarchy.
In these plots, $m_\Delta=1 \tev$, $m_S=7\tev$ and $|(Y_S)_{33}| =0.0097$.  The dashed lines represent the current experimental limits at 90\%C.L. }
\label{fig:NH_LFV}
\end{figure}
\end{center}

The correlations between these cLFV processes are displayed in Fig.\ref{fig:IH_LFV}(Fig,\ref{fig:NH_LFV}) for IH(NH).
The sign of $(Y_S)_{33}$ is responsible for the two  prominent clusters in each scatter plot.
However, the origin of the notable difference is mere technical and it can be traced back to  Eq.(\ref{eq:YL_leading}):
The two terms in the numerator of $(Y_L)^{(0)}_{i2}$ have compatible magnitudes.
So when the sign of $(Y_S)_{33}$ is right,  the two terms almost cancel out with each other  yielding a relatively small $|(Y_L)^{(0)}_{i2}|$.
The opposite happens when the sign of $(Y_S)_{33}$ is wrong.

All these plots have $|(Y_S)_{33}|$ fixed at its maximally allowed value $0.0097$.
From here, other configurations can be obtained by simply scaling down $Y_S$ by $Y_S \rightarrow \lambda^{-2} Y_S$  ( with $\lambda>1$ ). In response,  all the LFV processes branching ratios move up as $\lambda^4$\footnote{During the scaling, one needs to recheck the configuration is still phenomenologically viable. }. In some plots, the dashed arrows are put in to guide the reader's eyes and show the drifting direction of the
 branching ratios during the  re-scaling. As the $Y_S$ is dialed down, all the  points of ${\cal B}^Z_{ll'}$ go up along the indicated direction until the $B(\mu\rightarrow e\gamma)$ hits the current experimental limit. Interestingly, we can predict the upper limits on ${\cal B}^Z_{ll'}$, ranging from $10^{-11}$ to $10^{-9}$, for the $Y_R=0$ case.
We stress that these upper limits are tied  with the $Y_R=0$ assumption; they could be much larger if $Y_R\neq 0$.
 This part of parameter space of cZBM could be probed at the planned TeraZ collider where about $10^{12}$ Z bosons will be produced per year with a few $ab^{-1}$ luminosity\cite{TLEP,CEPC}. Moreover, this particular assumption will be ruled out if any excess was measured in the future experiment.
On the other hand, the lower limits on these LFV processes are rather robust and insensitive to the $Y_R=0$ assumption.
Similarly, interesting upper and lower bounds on ${\cal B}(\ell \rightarrow \ell' \gamma)$ and the CP asymmetries $\eta_{\ell \ell'}$ can be predicted in cZBM, see Tab.\ref{tab:range_Z_eta}. Note that the upper bounds on all three ${\cal B}(\ell \rightarrow \ell' \gamma)$ are just below the current experimental limits. Any  improvement in these measurements will cut across the interesting parameter space  of cZBM.
On the other hand, the cZBM with democratic $Y_S$ and $m_\Delta\sim{\cal O}(\tev)$ can be falsified if no such cLFV processes had been detected above the predicted lower bounds in the future experiments.
\begin{table*}[h]
\begin{center}
\begin{tabular}{c|c|c}
\hline\hline
										   &lower bounds  &  upper bounds (for $Y_R=0$)   \\
\hline
$\mathcal{B}( \mu\rightarrow e\gamma ) $ 	 & $3.05 \times 10^{-16}$ ($3.98 \times 10^{-18}$) 	& $ 5.7(5.7) \times 10^{-13} $  \\
$\mathcal{B}( \tau\rightarrow  e \gamma ) $  & $3.16 \times 10^{-16}$ ($2.03 \times 10^{-18}$)	& $ 2.3 (0.51)\times 10^{-9} $ \\
$\mathcal{B}( \tau \rightarrow \mu \gamma ) $ & $4.67 \times 10^{-17}$ ($1.68 \times 10^{-16}$)	& $ 3.4 (2.8) \times 10^{-8} $ \\
\hline\hline
$\mathcal{B}^Z_{e \mu} $   	  	 & $2.5 \times 10^{-16}$ ($4.9 \times 10^{-14}$)   & $2.2  (8.7) \times 10^{-11}$  \\
$\mathcal{B}^Z_{e \tau} 	$ 	 & $2.9 \times 10^{-16}$ ($4.6 \times 10^{-14}$)	& $ 3.6 (1.0 ) \times 10^{-10} $  \\
$\mathcal{B}^Z_{\mu \tau} $	  	 & $2.5 \times 10^{-14}$ ($7.8 \times 10^{-15}$)	& $ 5.5 (4.5 )\times 10^{-9} $     \\
\hline\hline
$\eta_{\mu e} $							     & 	${}^{+.68}_{-.67}({}^{+2.1}_{-.97}) \times 10^{-13}$						 &$ {}^{+2.6}_{-5.4} ({}^{+9.3}_{-8.1}) \times 10^{-13} $ \\
$\eta_{\tau e} 	$						     & 	${}^{+2.4}_{-.20}({}^{+.20}_{-1.2}) \times 10^{-12}$						 &$ {}^{+2.3}_{-.56} ({}^{+.22}_{-.10}) \times 10^{-11} $ \\
$\eta_{\tau \mu}$ 						     & 	${}^{+2.3}_{-.78}({}^{+1.3}_{-1.3}) \times 10^{-11}$				     	 & $ {}^{+3.7}_{-8.1} ({}^{+3.0}_{-3.1}) \times 10^{-11}$\\
\hline\hline
\end{tabular}
\end{center}
\caption{Range of $\mathcal{B}(\ell \to \ell' \gamma)$, $\mathcal{B}( Z \rightarrow \ell \ell' )$, and $\eta_{\ell \ell'}$  for $m_\Delta =1 \tev$ and $m_S=7\tev$. The numbers( in the parentheses  ) are for NH(IH) neutrino masses.
The lower bounds are the lowest values found in the numerical search  with $|(Y_S)_{33}| = 0.0097$.
The upper bounds are found by rescaling $Y_S$, see text.  Note that the sign for $\eta_{\ell \ell'}$ could be either ways.
For a different leptoquark/di-quark mass, all the values should be multiplied by a factor of $(1\tev\cdot m_S/7 m_\Delta^2)^2$.
}
\label{tab:range_Z_eta}
\end{table*}

\begin{center}
\begin{figure}[h]
  \subfigure[]{\includegraphics[width=0.3\textwidth]{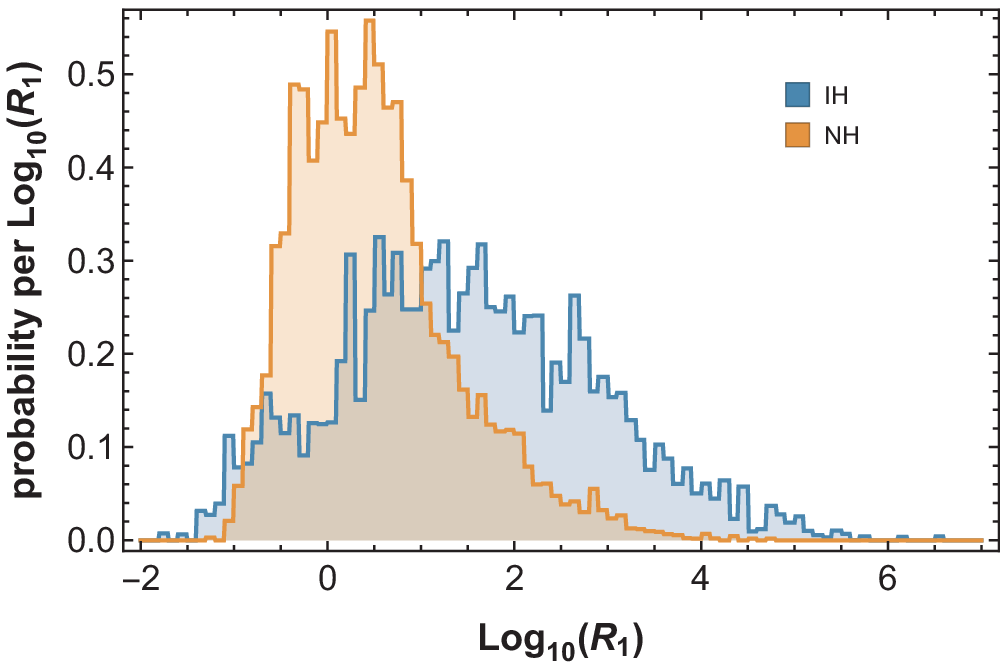}}
  \subfigure[]{\includegraphics[width=0.3\textwidth]{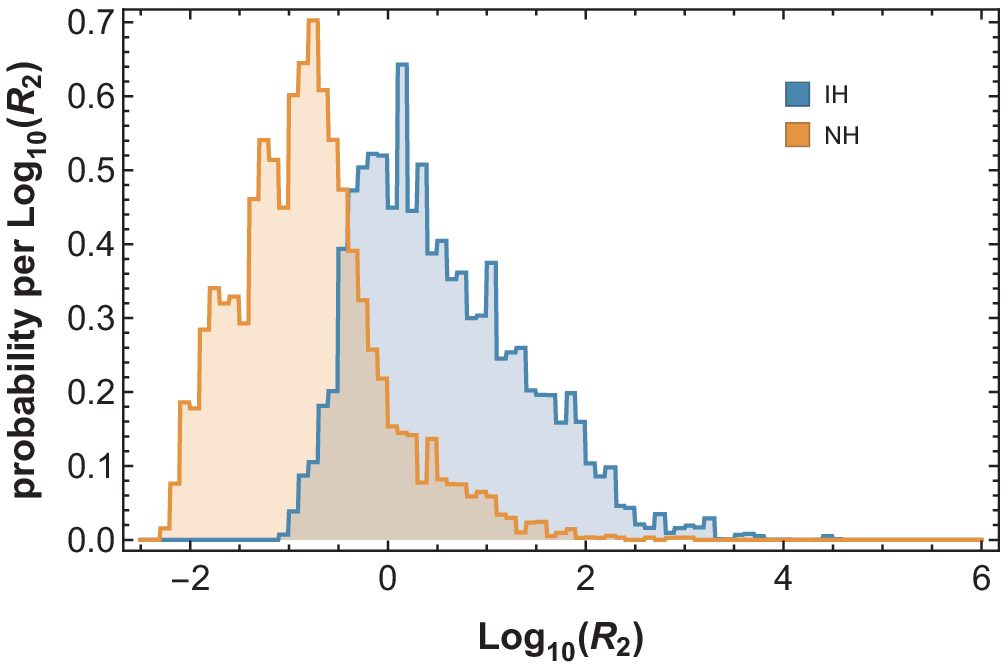}}
  \subfigure[]{\includegraphics[width=0.3\textwidth]{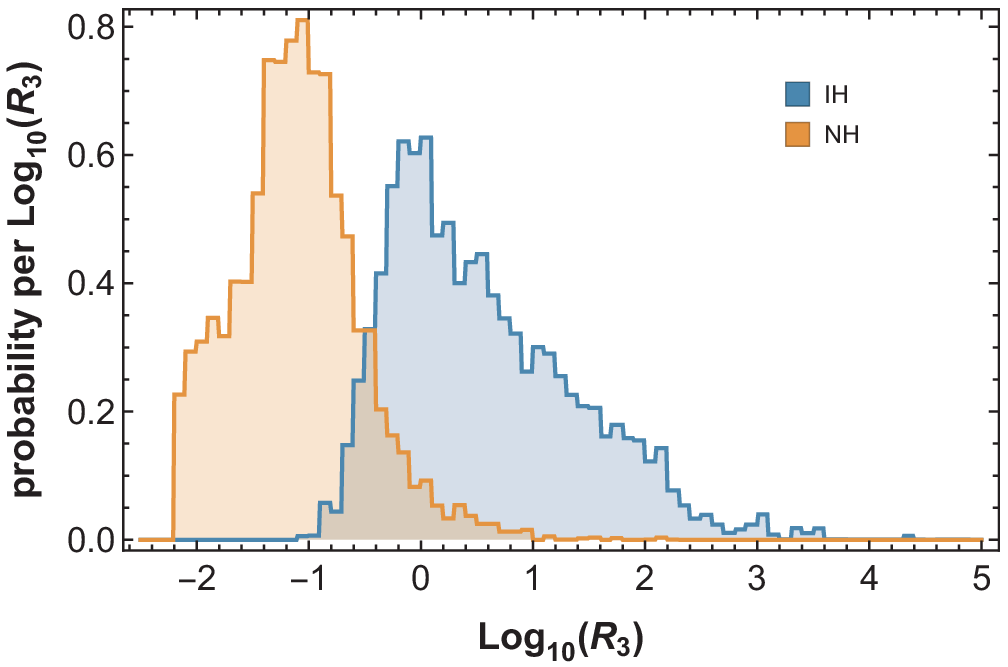}}\\
  \subfigure[]{\includegraphics[width=0.3\textwidth]{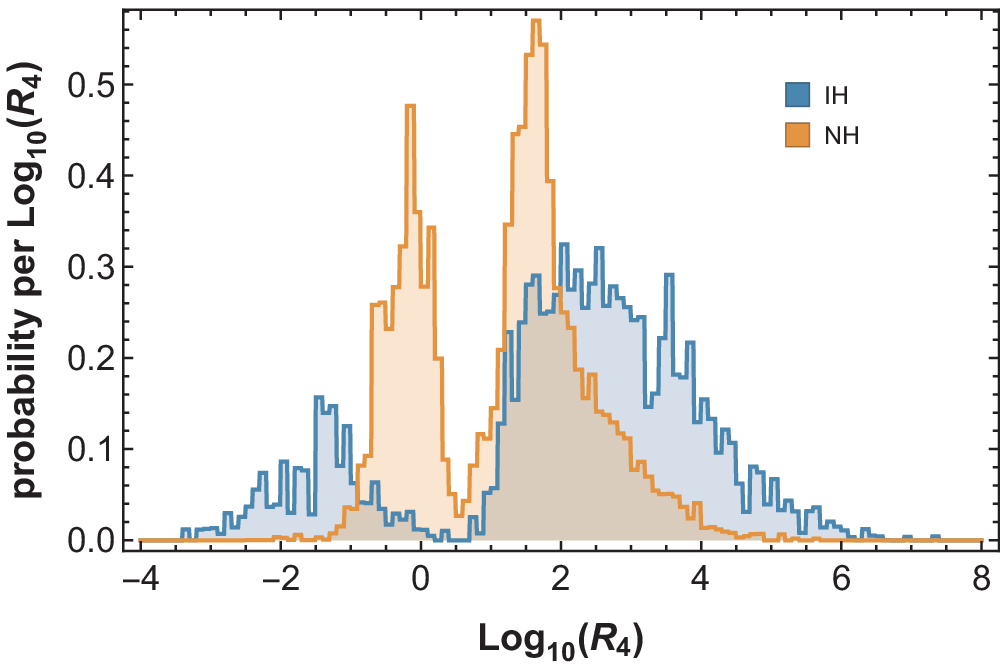}}
  \subfigure[]{\includegraphics[width=0.3\textwidth]{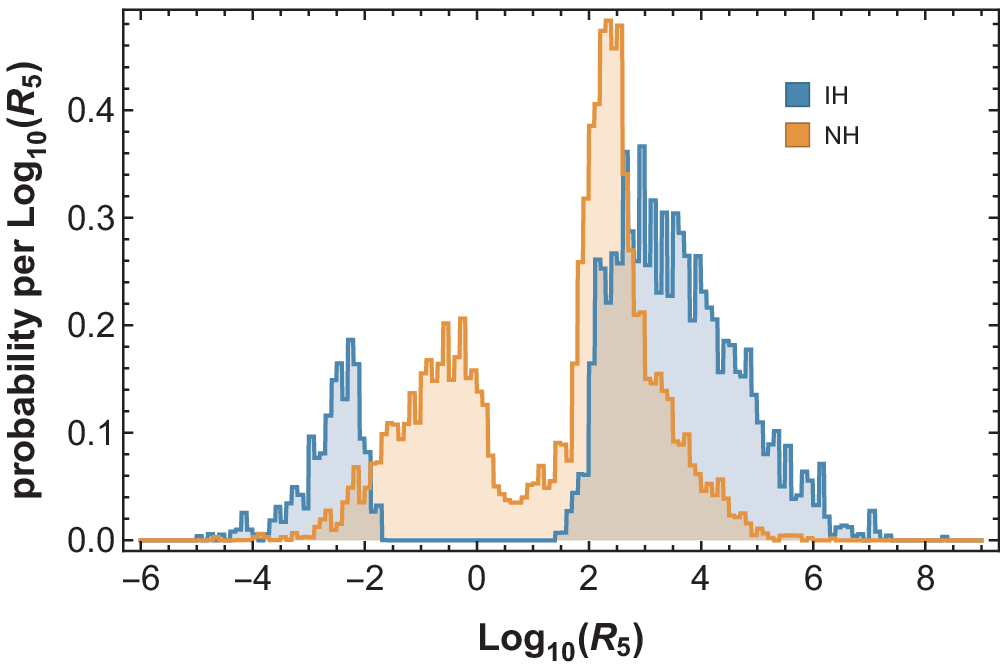}}
  \subfigure[]{\includegraphics[width=0.3\textwidth]{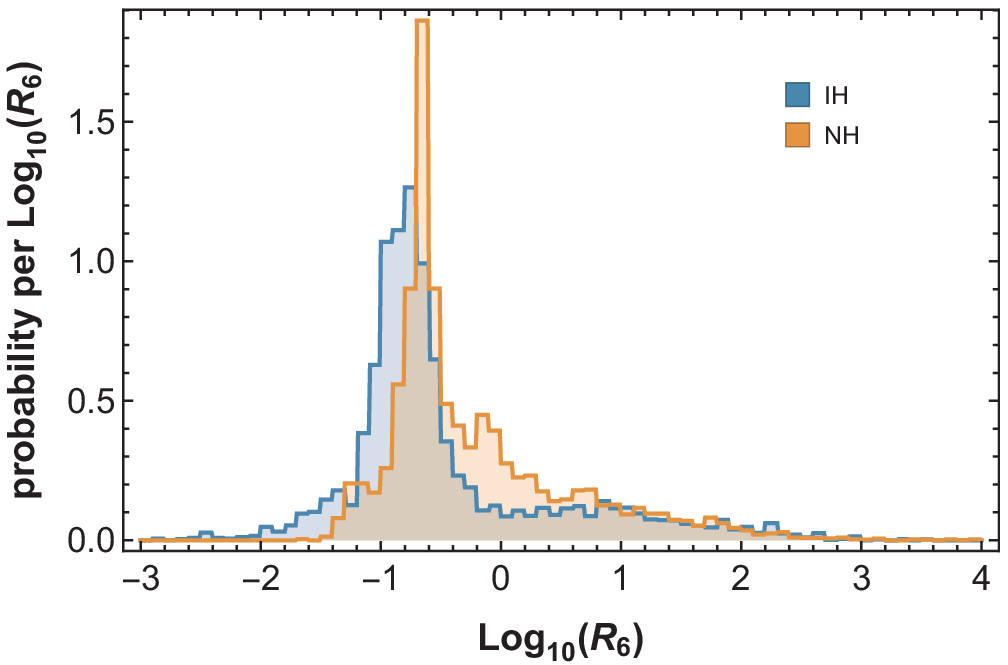}}\\
    \subfigure[]{\includegraphics[width=0.3\textwidth]{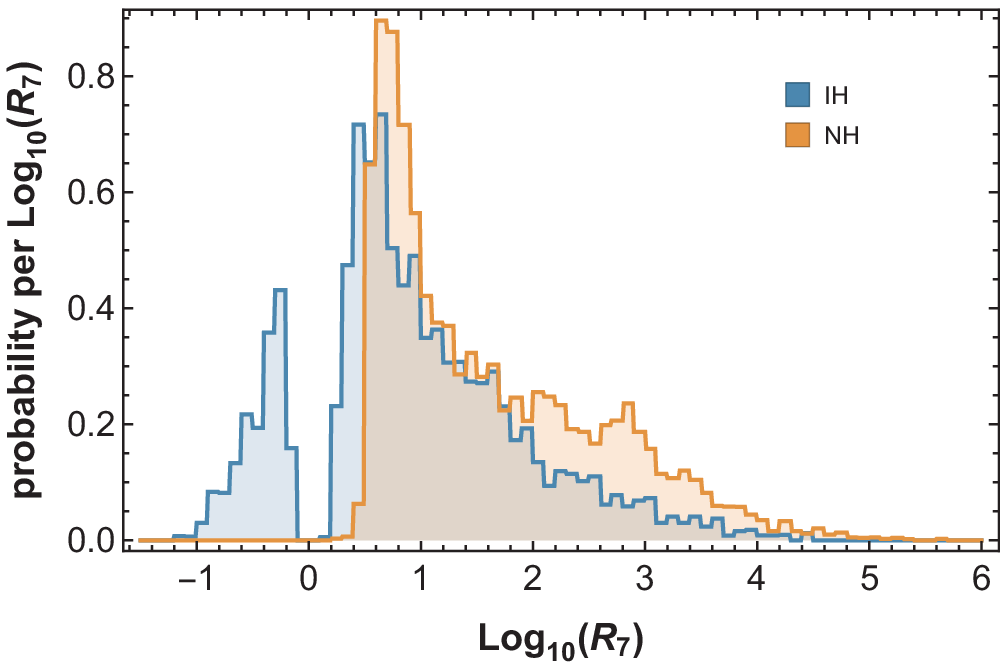}}
  \subfigure[]{\includegraphics[width=0.3\textwidth]{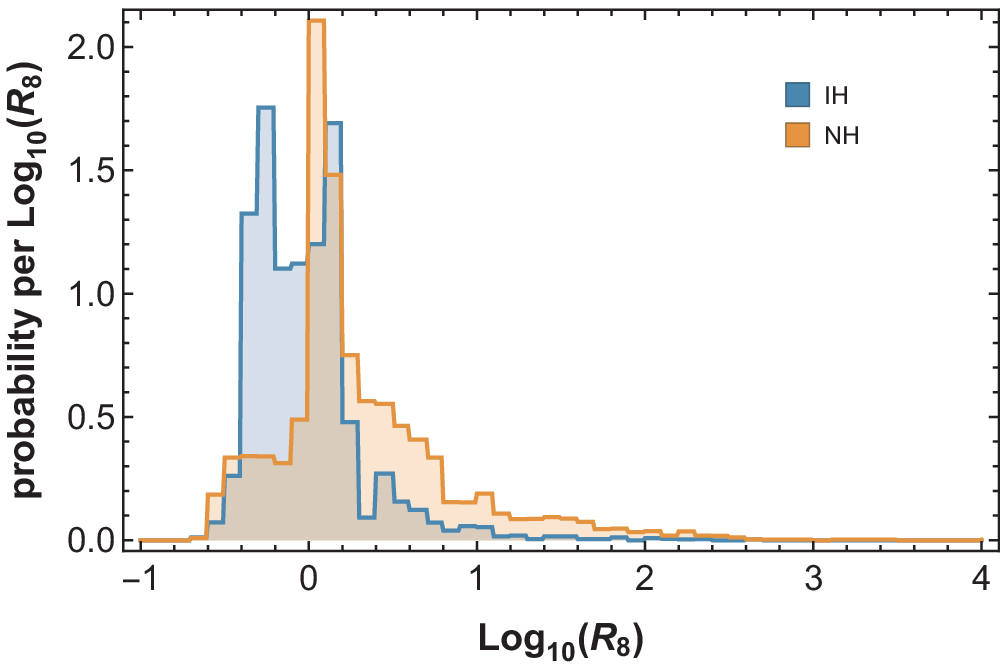}}
  \subfigure[]{\includegraphics[width=0.3\textwidth]{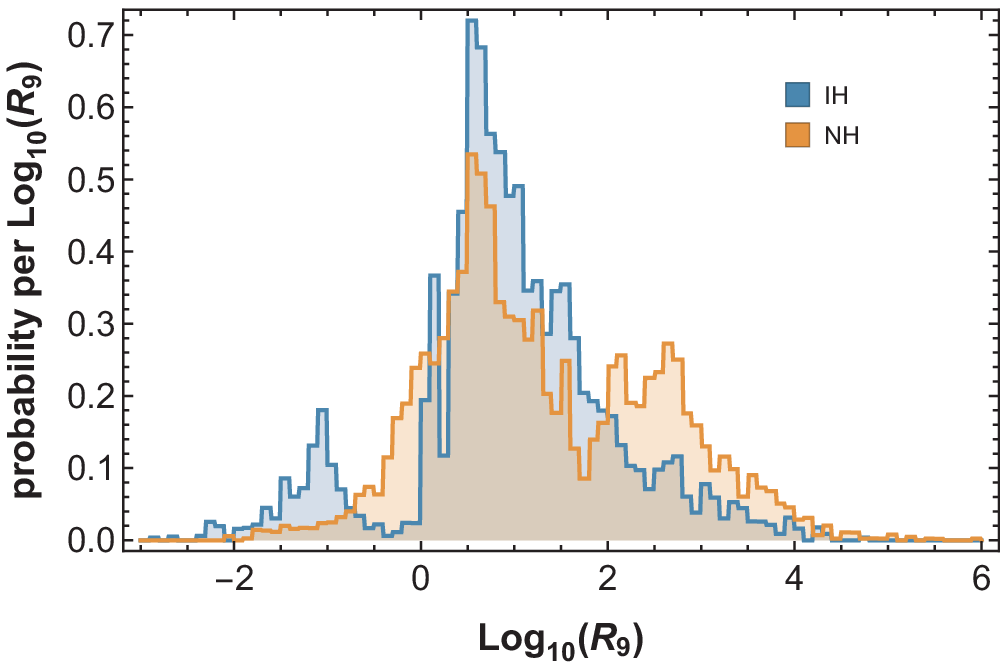}}\\
\caption{Side by side comparison of the double ratios in NH and IH. See Tab.\ref{tab:sweetspots} for the double ratio definitions. }
\label{fig:double_R}
\end{figure}
\end{center}

Note that the double ratio of any pair of cLFV process branching ratios is invariant under the $Y_S$ re-scaling and independent of $m_\Delta$. Our numerical also has concrete
predictions for $Y_R=0$ and these double ratios  depend on the neutrino mass pattern, see Fig.\ref{fig:double_R}. Therefore they could provide an intriguing mean to determine the neutrino mass hierarchy.
In particular, the neutrino mass hierarchy can be unambiguously  determined if the measured values fell into any of the decisive windows listed in Tab.\ref{tab:sweetspots}. Most of these interesting double ratio windows are plagued by either small cLFV branching ratios or very limited parameter space. However, $R_5\equiv {\cal B}(\tau\rightarrow \mu \gamma )/ {\cal B}(\tau\rightarrow e \gamma )$ and $R_7\equiv {\cal B}^Z_{\mu\tau}/ {\cal B}^Z_{\mu e}$ look quite promising. In the cZBM, if  $R_5$ is measured in the future rare tau decay experiment to be within $0.03$ and $30$, the neutrino masses are of NH. If $R_7<1.0$ is measured in the future $Z$-factory, the neutrino masses are of IH in the cZBM. Even in the worst scenario that none of the measured double ratios overlap with these stated windows, one could still tell which neutrino mass hierarchy is more probable by simple statistics and probability theory. For example, if both $R_4$ and $R_5$ were measured to be $\sim 10^3$, then the IH is roughly 4 times more probable than the NH in the cZBM.
The above discussion clearly demonstrates that  the neutrino oscillation experiments and the cLFV measurements are complimentary to one another to better understand the origin of the neutrino masses.

\begin{table*}[h]
\begin{center}
\begin{tabular}{c|c|c}
\hline\hline
Double Ratio & IH  & NH \\
\hline
$R_1 \equiv {\cal B}^Z_{\mu\tau}/ {\cal B}(\mu\rightarrow e \gamma )$ & $R_1>10^4$ or $R_1<0.1$ & N.A.  \\
$R_2 \equiv {\cal B}^Z_{e \tau}/ {\cal B}(\mu\rightarrow e \gamma )$ & $R_2>10^3$ & $R_2<0.1$ \\
$R_3 \equiv {\cal B}^Z_{e \mu}/ {\cal B}(\mu\rightarrow e \gamma )$& $R_3>10^2$ & $R_3<0.1$ \\
$R_4 \equiv {\cal B}(\tau\rightarrow \mu \gamma )/ {\cal B}(\mu\rightarrow e \gamma )$ & $R_4>10^6$ & $R_4<0.003$\\
$R_5 \equiv {\cal B}(\tau\rightarrow \mu \gamma )/ {\cal B}(\tau\rightarrow e \gamma )$ & N.A. &  $0.03 <R_5 <30$ \\
$R_6 \equiv {\cal B}(\tau\rightarrow e \gamma )/ {\cal B}(\mu\rightarrow e \gamma )$ & $R_6<0.03$ & N.A. \\
$R_7 \equiv {\cal B}^Z_{\mu\tau}/ {\cal B}^Z_{\mu e}$ & $R_7<1.0 $  &  $R_7>3\times 10^4$  \\
$R_8 \equiv {\cal B}^Z_{e \tau}/ {\cal B}^Z_{e \mu }$ & N.A. & $R_8>10^2$ \\
$R_9 \equiv {\cal B}^Z_{\tau \mu}/{\cal B}^Z_{\tau e}$& $R_9<0.01$ & $R_9>3\times 10^4$ \\
\hline\hline
\end{tabular}
\end{center}
\caption{ The definitions of the cLFV double ratios and the ranges which can be used to determine neutrino mass hierarchy.
}
\label{tab:sweetspots}
\end{table*}

\subsubsection{Leptoquark decay branching ratios}
First, some shorthand notations are introduced:
\beq
\Gamma^{\ell^-_i}_\Delta \equiv \sum_{j} \Gamma(\Delta \rightarrow \ell^-_i u_j)\,,
\Gamma^{\nu_i}_\Delta \equiv \sum_{j} \Gamma(\Delta \rightarrow \nu_i d_j)\,,
\eeq
where the quark flavors are summed over.
For $Y_R=0$, the $SU(2)_L$ symmetry ensures that $\Gamma^{\ell^-_i}_\Delta =\Gamma^{\nu_i}_\Delta\propto |\sum_j(Y_L)_{\ell_i j}|^2$.
This corresponds to the $\beta=1/2$ case that 50\% of the leptoquark decays into a neutrino and a down-type quark.
Since the neutrino is hard to be tracked in the detector, we  focus on the decay channels with a high energy charged lepton as the primordial final state and define
\beq
B^\Delta_{\ell_i} \equiv { \Gamma^{\ell^-_i}_\Delta \over  \Gamma^{e^-}_\Delta+\Gamma^{\mu^-}_\Delta+\Gamma^{\tau^-}_\Delta}\,.
\eeq
The above defined quantity is clearly independent of  $m_\Delta$ and $Y_S$ re-scaling.

\begin{center}
\begin{figure}[h]
 \subfigure[]{  \includegraphics[width=0.44\textwidth]{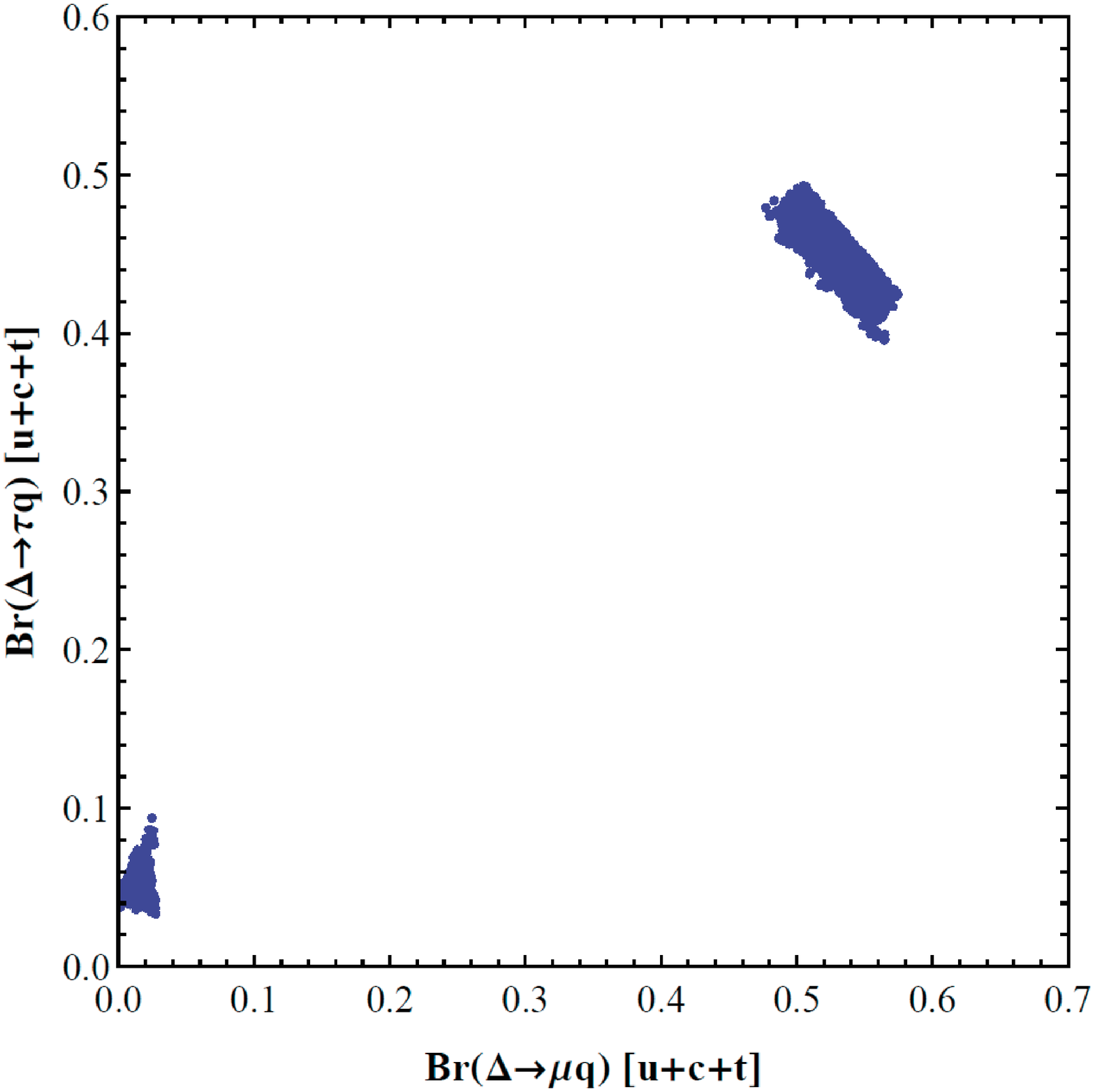}}\,
  \subfigure[]{ \includegraphics[width=0.43\textwidth]{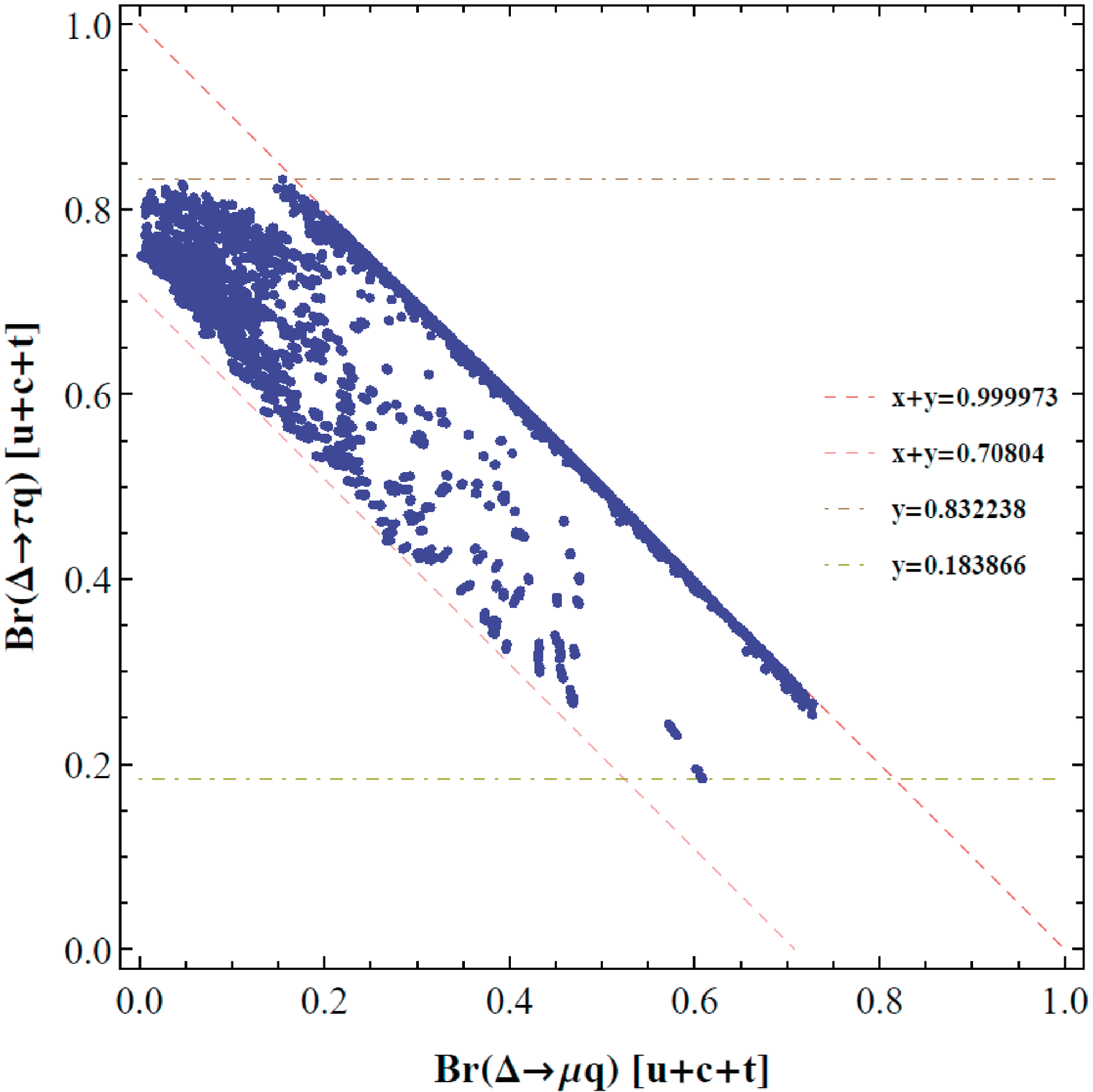}}
\caption{leptoquark decay branching ratios for (a) IH, (b) NH. }
\label{fig:LQ_BR}
\end{figure}
\end{center}
 The decay branching ratios for leptoquark  from our numerical study are displayed  in Fig.\ref{fig:LQ_BR}.
It can be clearly seen in  Fig.\ref{fig:LQ_BR} that, depending on the neutrino mass hierarchy, there are special patterns in the leptoquark decay branching ratios. Roughly speaking, for the IH case, the leptoquark decays are either (1) $B^\Delta_{e}\sim 1.0$ or (2) $B^\Delta_{\mu}\sim 55\%$ and $B^\Delta_{\tau}\sim 45\%$. On the other hand, for the NH case,  the $B^\Delta_{\mu}$ and $B^\Delta_{\tau}$  are concentrated in the region roughly enclosed by  $0.7 \lesssim  B^\Delta_{\mu}+B^\Delta_{\tau}\lesssim  1.0$ and $ 0.2\lesssim B^\Delta_{\tau}\lesssim 0.8$. In other words, $ B^\Delta_{e} \lesssim 0.3$ if the neutrino masses are in the NH.

Surprisingly, our numerical study has not found any configuration which has either $B^\Delta_{\mu}\sim 100\%$ or $B^\Delta_{\tau}\sim 100\%$.
Dictated by the neutrino oscillation data, the model predicts that the leptoquark can NOT decays  purely into the 2nd or the 3rd generation charged leptons. These concrete branching ratios could be used to provide the new benchmark leptoquark mass limits with a better motivation.

\section{Conclusion}
We have studied the cZBM which exploits a scalar leptoquark $\Delta(3,1,-1/3)$ and a scalar di-quark $S(6,1,-2/3)$  to generate neutrino masses at the 2-loop level. The neutrino mass matrix element $M^\nu_{ij}$ ( i,j=1,2,3) is proportional to the product of $(Y_L)_{ik} m^d_k (Y_S^\dag)_{k k'}m^d_{k'} (Y_L^T)_{k' j}$, see Eq.(\ref{eq:nu_mass_elements}). The Yukawa couplings $Y_L$ and $Y_S$  are a priori unknown and arbitrary. To proceed, we have adopted a modest working assumption that all six $|Y_S|$ are of the same order.  Then the $Y_L$ can be iteratively determined owing to the fact that $m_\Delta, m_S\gg m_b\gg m_s\gg m_d$. Moreover,  the mass of the lightest neutrino is of order $10^{-5}$ eV and the model disfavors the case of nearly degenerate neutrinos.  The tree-level flavor violating processes will be inevitably mediated by $\Delta$ or $S$ with the realistic $Y_L$ which accommodates the neutrino data. Due to the different chiral structures, the contributions to the flavor violating processes  from $Y_L$ and $Y_R$ do not interfere with each other at the tree-level.  We have considered the case that $Y_R=0$ to minimized the tree-level flavor violating processes( and expect the same to happen for the loop induced cLFV). We also have argued that $Y_R=0$ is actually favored by the fact that there is no  electron EDM has been observed yet.
A comprehensive numerical study has been performed to look for the realistic $Y_L$ and $Y_S$ configurations which pass all the known experimental constraints on the flavor violating processes.  The viable configurations were collected and have been used to calculate the resulting 1-loop charged lepton flavor violating $Z \rightarrow \overline{l} l'$ and $l \rightarrow l' \gamma$.
Some of the realistic configurations could be probed in the forthcoming cLFV experiments.  Interesting and robust lower bounds have been found for these cLFV, see Tab.\ref{tab:range_Z_eta}. Moreover, the neutrino mass hierarchy could be determined if the measured cLFV double ratio(s)  is/are in some specific range(s), see Tab.\ref{tab:sweetspots}.
For $Y_R=0$, $\Delta$ has $50\%$ of chance decaying into a charged lepton and an up-type quark. Specific ratios
$\sum_j B(\Delta\rightarrow l_i u_j)$ for each generation charged lepton $l_i$ have been predicted in this model.
Given the potential link between the neutrino masses generation and $\Delta$, it seems well-motivated  using the predicted leptoquark branching ratios as a benchmark scenario for the future scalar leptoquark search limits.

\begin{acknowledgments}
 WFC is supported by the Taiwan Minister of Science and Technology under
Grant Nos. 102-2112-M-007-014-MY3 and 105-2112-M-007-029.
FX is supported partially by NSFC (National Natural Science Foundation of China ) under Grant No. 11605076,
as well as the FRFCU (Fundamental Research Funds for the Central Universities in China) under the Grant No. 21616309. FX especially acknowledges the hospitality of Institute of Physics,
Academia Sinica, at which part of the work was done.
\end{acknowledgments}
%%%%%%%%%%%%%%%%%%%%%%%%%%%%

%%%%%%%%%%%%%%%%%%%%%%%%%%%%%%%%%%%%%%%%%%%%%%%

\end{document}